\documentclass[onecolumn,10pt]{article}
\usepackage[dvips]{color}

\usepackage{epsfig}
 \usepackage[utf8]{inputenc}
\usepackage{amsmath}
\usepackage{amssymb}
\usepackage{txfonts}
\usepackage{bm}
\usepackage{graphicx}
\usepackage{dsfont}
\usepackage{xcolor}
\usepackage{multirow}
\usepackage{cite}
\usepackage{hyperref}
\usepackage{orcidlink}
\hypersetup{
 colorlinks,
 linkcolor={blue},
 citecolor={green!80!black},
 urlcolor={red}
}

\begin{document}

\title{\bf Study of Stable Dark Energy Stars in Ho$\Check{\text{r}}$ava-Lifshitz gravity}

\author{\bf{Krishna Pada Das  \orcidlink{0000-0003-4653-6006} $^1$}\thanks {krishnapada0019@gmail.com}~~~and~~
\bf{Ujjal Debnath  \orcidlink{0000-0002-2124-8908} $^1$} \thanks{ujjaldebnath@gmail.com}\\
$^1$Department of Mathematics, Indian Institute of Engineering\\
Science and Technology, Shibpur, Howrah-711 103, India.\\
}

\maketitle

\begin{abstract}

We study the structure and basic physical properties of non-rotating dark energy stars in Ho$\Check{\text{r}}$ava-Lifshitz (HL) gravity. The interior of propsed stellar structure is made of isotropic matter obeys extended Chaplygin gas EoS. The structure equations representing the state of hydrostatic equilibrium i.e., generalize TOV equation in HL gravity is numerically solved by using chosen realistic EoS. Next, we investigate the deviation of physical features of dark energy stars in HL gravity as compared with general relativity (GR). Such investigation is depicted by varying a parameter $\omega$, whereas for $\omega\rightarrow \infty$ HL coincide with GR. As a results, we find that necessary features of our stellar structure are significantly affected by $\omega$ in HL gravity specifically on the estimation of the maximum mass and corresponding predicted radius of the star. In conclusion, we can predict the existence of heavior massive dark energy stars in the context of HL gravity as compared with GR with not collapsing into a black hole. Moreover, we investigate the stability of our proposed stellar system. By integrating the modified perturbations equations in support of suitable boundary conditions at the center and the surface of the stellar object, we evaluate the frequencies and eigenfunctions corresponding to six lowest excited modes. Finally, we find that physically viable and stable dark energy stars can be successfully discussed in HL gravity by this study.

\end{abstract}

\section{Introduction}

It is no doubt that GR is the most successful theory of gravity for explaining astrophysical phenomenon of our current Universe including the motion of stellar objects and planets till now. But based on different observational data \cite{riess1998observational,perlmutter2003measuring,de2000flat,hanany2000maxima,ade2016planck,akrami2018planck,aghanim2020planck}, it is predicted that GR faces some problems to explain many astrophysical phenomenon in quantum nature as well as at strong gravitational field near compact relativistic objects like black holes, neutron stars and mysterious darkness such as dark energy and dark matter. Here it is important to mention that our current universe belong in an accelerating phase \cite{garnavich1998supernova,riess1998observational} and the main hidden source of energy that actually produces the negative energy repulsive to gravitation force, named as dark energy (DE). Moreover, it is observed that $95-96 \%$ of the the total content of our current universe is in the form of dark energy and dark matter, whereas the remaining $4-5 \%$ represented by baryonic matter \cite{peebles2003cosmological,padmanabhan2003cosmological}. Now, to addresses these issues several models regarding the origin and nature of dark energy as well as alternative theories of gravity have been introduced. Usually, DE defined as an negatively pressurized source of energy that violates the strong energy condition (SEC) i.e., $\rho + 3p <0$ with $\rho$ and $p$ are energy density and pressure respectively. In general, the EoS of DE is defined as $p=w\rho$ with the parameter $w<-1/3$. For the range $-1<w<-1/3$ DE referred as quintessence and for $w<-1$ ghost or phantom like DE. The most simplest form of DE is represented by Cosmological constant $\Lambda$. Again the existence, origin and exact nature of dark matter is still also an open issue. According to the refs.~\cite{munoz2004dark} approximately $85\%$ of the total mass of the Universe is in the form of dark matter.\\
Compact stars are defined as the final state of very massive stellar object, are extremely dense type objects \cite{shapiro1983physics}. The radius of such objects are of a few kilometers and with mass of a order of one solar mass like neutron stars \cite{lattimer2007neutron}, strange stars \cite{alcock1986strange,weber2005strange} and other class of these objects are remaining hypothetical type objects. 
Thus the reobservation of characters as well as predicted maximum mass for astrophysical objects within the framework of modified theories of gravity become a new platform in modern cosmological research \cite{wei2008distinguish,wang2008differentiating,bertschinger2008distinguishing,nojiri2005modified,joyce2016dark}. Among these, an outstanding prediction that dark energy may be exist in the interior of astrophysical objects has introduced by several researchers. In this connection, recently Mazur and Mottala \cite{mazur2023gravitational,mazur2004gravitational,mottola2010new,mazur2015surface} proposed an exact solution of the final state of compact objects so the called gravastars (or gravitational vacuum stars) with interior false vacuum region, filled by DE. Based on the physical structure of gravastar, two types of compact stellar objects like one made by dark energy and other one is made by a non-iteracting mixture of normal matter and dark energy \cite{harko2011two,rahmansyah2021tov}, called by dark energy stars have been predicted. Now, we want to concentrate some literature regarding physical characteristics of dark energy stars. Generalizing the concept of gravastar picture, Lobo explored several properties and also the stablity under  linear perturbation of dark energy stars \cite{lobo2006stable,lobo2007van}. An exact solution for describing relativistic stellar structure made by a mixture of ordinary matter and phantom like dark energy in different ratio has been constructed by Yazadjiev \cite{yazadjiev2011exact}. Different properties of dark energy stars with interior anisotropic pressure has been discussed by Chan et al., \cite{chan2009star}. Also, anisotropic DE stars with Tolman IV - type gravitational potential functions have been explored in the ref.~\cite{dayanandan2021modelling}. Ghezzi \cite{ghezzi2011anisotropic} explored the mass-radius relation for DE stars formed by isotropic ordinary matter and anisotropic type DE. Again Rahaman et al. \cite{rahaman2012singularity}, discribed a stable DE star configuration with cosmological constant $\Lambda$ type DE EoS ($w=-1$) satisfying the relation dark energy density is proportional to the ordinary matter density. Using this relation, the configuration of DE stars have also been explored by many researches \cite{bhar2015dark,das2015anisotropic,das2023dark,bhar2023dark,majeed2023some}. The significant effect of rainbow function on the configuration of dark energy stars has been explored in \cite{tudeshki2022dark,tudeshki2024effect}. In the context of  Vegh’s massive gravity, dark energy star has been studied in the ref.~\cite{tudeshki2023effect}. Some extra features like Radial pulsations, moment of inertia and tidal deformability of dark energy stars have been successfully discussed by Pretel \cite{pretel2023radial}.\\

Nowadays, an important challenge is the issue of ultraviolet (UV) divergence at extreme situation of high energies due to quantum effect at the early stage of the Universe or in the vicinity of a black hole. Commonly this problem is known as UV completion problem \cite{stelle1977renormalization}. In a series of papers \cite{hovrava2009membranes,hovrava2009quantum,hovrava2009spectral} Ho$\Check{\text{r}}$ava addressed the UV complete theory by sacrificing local Lorentz symmetry, named as Ho$\Check{\text{r}}$ava-Lifschitz (HL) gravity. Here the Lorentz symmetry is described by a Lifshitz-type rescaling process \cite{minamitsuji2010classification}
\begin{center}
$t\rightarrow b^{z} t$~~~~\text{and}~~~~$x^{i}\rightarrow b x^{i}$~~~~\text{with}~~~~$i=1, 2, 3,...,d$.
\end{center}
Here d denotes the dimension of spacetime, z is the dynamical exponent for scaling and b is an arbitrary constant. In 4-dimensional spacetime, HL gravity is power counting renormalizable that provide $z\ge 3$ \cite{hovrava2009membranes,visser2009lorentz}. In the present study, we take $z=3$.
Further by including higher order curvature term, describing by a new parameter $\omega$ in the original gravitational action of HL gravity, an another theory so the called deformed HL gravity has been introduced \cite{hovrava2009membranes,park2009black}. Here an important fact that GR can be recovered at the infrared (IR) scale by approximating HL gravity.  Kehagias and Sfetsos (KS) evaluated a solution for asymptotically flat and static spherically symmetric geometry in HL gravity that approaches the Schwarzschild black hole (SBH) at the IR limit \cite{kehagias2009black}. Utilizing KS solution, the constraints on HL gravity have been studied in the refs.~\cite{liu2011solar,horvath2011constraining,iorio2011hovrava}. In a lot of literature, dark energy, dark matter, various black hole solutions and thermodynamics of theme have already been studied in HL gravity by several authors \cite{saridakis2010hovrava,park2009black,mukohyama2009dark,cai2009topological,myung2010thermodynamics,mann2009lifshitz,castillo2009entropy,peng2010hawking,colgain2009dyonic,lee2010extremal,kim2009surplus,koutsoumbas2010black,majhi2010hawking}.\\

Now, we will concentrate on the exploration of compact stellar structure such as white dwarfs (WD) and neutron stars (NS) in the context of various modified theories of gravity (MTGs) including HL gravity. Basically, the configuration of a static and spherically symmetric compact stellar structure are explored through the TOV equation \cite{tolman1939static,oppenheimer1939massive} in GR and modified version of its in MTGs. One of the vital feature of TOV equation is that by solving this equation, one can easily obtain the mass-radius relation and consequently, the maximum mass and the total radius of a compact star. Also, the stiffness of proposed equation of state (EoS), representing the nature of interior matter portion of the star, can easily describe by solving TOV equation. Various properties of a NS have been explored in the context of different MTGs in several literature \cite{astashenok2013further,astashenok2014maximal,astashenok2015magnetic,astashenok2015extreme,kim2014physics,harko2013structure,prasetyo2018neutron}. By solving of Maxwell equations, the stellar magnetic configuration for a spherically magnetized star in HL gravity as well as in $f(R)$ gravity are discussed in \cite{hakimov2013magnetic}. Static and spherically symmetric stars and black hole are studied in HL gravity with  the projectability condition in \cite{greenwald2010black}. Also, with the projectability condition, gravitational collapse of a spherical fluid has been studied in \cite{greenwald2013gravitational}. In the ref.~\cite{kim2021neutron}, authors have successfully presented aspects of NSs specially mass-radius relation in deformed HL gravity by constructing modified TOV equation. As a result, authors have found that a NS with larger radius and heavier mass but not collapsing into a black hole may be presented in HL gravity compared with GR. Moreover, authors have compared the results with the observed maximum mass of NSs $\sim 2M_{\odot}$ in GR \cite{demorest2010two,linares2018peering}.\\

In the current study, our main motivation is to investigate the possible formation of a static dark energy star in deformed HL gravity. 
\textbf{It is worthwhile to mention that accelerating expansion nature of our current universe suggests the existence of a cosmic fluid, which is negative pressurized in nature and anti-gravitational, called by dark energy (DE) \cite{peebles2003cosmological}. Again, current modern data based on several observational tests seem that approximately 70\% of the universe is made of DE. So, a special consideration coming up the possible existence of DE in the interior of other structure such as compact objects. On the other side, the development of alternative candidate of black hole to avoid central singularity and event horizon become a new era of astroparticle research. In this connection, different hypothetical stellar structures like gravastars \cite{mazur2023gravitational,mazur2004gravitational}, non-singular black holes \cite{dymnikova1992vacuum}, false vacuum bubbles \cite{coleman1980gravitational} are already discovered by using different type of dark energy models in the stellar structure. By motivated these objects, Chapline first presented the concept of dark energy stars (DES). The idea behind this is explored as at a critical surface, the falling matter is converted into vacuum energy, which is much larger than the cosmic vacuum energy, creating a negative pressure to act against gravity \cite{chapline2005dark}. In advantage, any singularity does not occur in the interior of the star. Moreover, recent observational tests suggest that DE is very homogeneous and not very dense. So, in the stellar structure of the compact objects the equation od state of DE can be used as a definition of the interior fluid.
}
Without loss of generality, we assume the interior matter portion of proposed stellar object satisfy more extended Chaplygin gas EoS. Here we adopt the modified TOV equation in HL gravity from thr ref.~\cite{kim2021neutron} corresponding to a static and spherically symmetric line element. Although there are no observed DE star yet, however we want to emphasize on the mass-radius relation of stable DE stars and estimate the maximum mass and total radius of them in HL gravity. Moreover, we want to present the deviation of obtained results from GR in HL gravity. The rest portion of our current discussion are arranged as follows: in the next section, we have presented a brief review of structural and hydrostatic equilibrium equations for a static, spherically symmetric stellar structure in HL gravity. The description of the EoS specifically extended Chaplygin gas EoS is discussed in the Section~\ref{sec3}. The Section~\ref{sec4} is devoted to explore numerically obtained results for our considered dark energy star. Also, a detail graphical discussion is presented in that section. Finally, in Section~\ref{sec5}, we have reported our main conclusions of our whole study for dark energy stars in HL gravity. Note that we adopt the mostly used metric signature $(-~+~+~+)$ in this study.

\section{Hydrostatic Equilibrium Equation in HL Gravity}\label{sec2}
The action of HL gravity can be given as \cite{gao2010cosmological,park2009black}
\begin{eqnarray}\label{1}
\mathcal{I}_{HL}=\int dt d^{3}x \sqrt{g}N\Bigg[ 2 \frac{\left(\mathcal{K}_{ij}\mathcal{K}^{ij} - \lambda \mathcal{K}^{2}\right)}{\kappa^{2}} - \frac{\kappa^{2}}{2 \zeta^{4}}\left(\mathcal{C}_{ij} - \frac{\mu \zeta^{2}}{2}\mathcal{R}_{ij}\right)\left(\mathcal{C}^{ij} - \frac{\mu \zeta^{2}}{2}\mathcal{R}^{ij}\right) + \frac{\kappa^{2}\mu^{2}(4\lambda - 1)}{32(3\lambda - 1)}\left(\mathcal{R}^{2} + \frac{4(\omega - \Lambda_{W})}{4\lambda - 1}\mathcal{R} + \frac{12\Lambda_{W}^{2}}{4\lambda - 1}\right)\Bigg].
\end{eqnarray}
Here it is noted that the above action (\ref{1}) is prepared by an anisotropic scaling for timr and space as $b \rightarrow b^{3}t$ and $x^{i} \rightarrow b x^{i}$. Also, the tensor term $\mathcal{K}_{ij}$ represents an extrinsic curvature, given by \cite{gao2010cosmological,park2009black}
\begin{eqnarray}\label{2}
\mathcal{K}_{ij}=\frac{\Dot{g_{ij}} - \nabla_{i}N_{j} - \nabla_{j}N_{i}}{2N},
\end{eqnarray}
with $N$ is a Lapse function and $N^{i}$ is a shift vector. The spatial metric tensor id denoted by $g_{ij}$ and the tensor term $\Dot{g_{ij}}$ means $\frac{\partial g_{ij}}{\partial t}$. Again the tensor term $\mathcal{C}^{ij}$ named as Cotton-York tensor, defined by \cite{gao2010cosmological,park2009black}
\begin{eqnarray}\label{3}
\mathcal{C}^{ij}=\epsilon^{ikl}\nabla_{k}\left(\mathcal{R}^{j}_{l} - \frac{\delta^{j}_{l}\mathcal{R}}{4}\right),
\end{eqnarray}
where $\mathcal{R}_{ij}$ denotes the usual spatial Ricci tensor and $\mathcal{R}$ is the scalar curvature. \\
$\lambda$ is an additional dimensionless coupling constant and $\kappa^{2}$ is another coupling constant related to the Newtonian gravitational constant $G_{N}$. Also, three dimensionless constants like $\Lambda_{W}$, $\mu$ and $\zeta$, are stemmed from topologically massive gravity action \cite{deser1982three}. \\
For $\lambda = 1$ the Einstein-Hilbert action can be recovered in IR limit by identifying the fundamental constants with $c=\frac{\kappa^{2}}{4}\left[\frac{\mu^{2}(\omega - \Lambda_{W})}{3\lambda -1}\right]^{1/2}$, $G_{N}=\frac{\kappa^{2}c^{2}}{32\pi}$, and $\Lambda=-\frac{3}{2}\frac{\Lambda_{W}^{2}}{(\omega - \Lambda_{W})}$ are the respective representation of speed of light, Newtonian gravitational constant and cosmological constant. \\
So, the action of HL gravity (\ref{1}) consists six parameters like $\kappa, \mu, \zeta, \Lambda_{W}, \omega ~\text{and}~\lambda$. For simplicity, we have set $\lambda =1$ and $\Lambda_{W}=0$. Again, $\kappa$, $\mu$ are hidden in the physical constants related to $G_{N}$ and $c$ in the IR regime and $\zeta$ does not contributed to this non rotating configuration. Thus, the parameter $\omega$ is the only free parameter of our proposed HL gravity. Therefore, the effect of $\omega$ on the configuration of our proposed stellar structure will be shown in the later section.

Now, we consider a line element that represents the interior geometry of a static, spherically symmetric object as \cite{kim2021neutron}
\begin{eqnarray}\label{4}
ds^{2}= - e^{2\Phi(r)}c^{2}dt^{2} + \left[1 - f(r)\right]^{-1}dr^{2} + r^{2}\left(d\theta^{2} + \sin^{2}\theta d\phi^{2}\right)
\end{eqnarray}
Next we assume that the interior part of our proposed stellar object made by a perfect fluid, represents by the following energy-momentum tensor
\begin{eqnarray}\label{5}
\mathcal{T}_{\mu\nu}=(\rho + p)\mathcal{U}_{\mu}\mathcal{U}_{\nu} + p g_{\mu\nu}.
\end{eqnarray}
where $\rho = \rho(r)$ is the energy density of matter, $p=p(r)$ ia the pressure, and $\mathcal{U}_{\mu} = (1, 0, 0, 0)$ is the fluid 4-velocity vector.\\
Now, the equation of motion in HL gravity corresponding to the action (\ref{1}) and the energy-momentum tensor (\ref{5}) may be given by the following equations \cite{kim2021neutron}
\begin{eqnarray}\label{6}
\frac{1}{r^{2}\omega}\left[2r\omega f + \frac{f^{2}}{r}\right]'=\frac{16\pi G_{N}}{c^{2}}\rho,
\end{eqnarray}
\begin{eqnarray}\label{7}
\frac{1}{r^{4}\omega}\left[f(f - 2 r^{2}\omega) + 4r(1 - f)(f + r^{2}\omega)\Phi'\right]=\frac{16 \pi G_{N}}{c^{4}}p,
\end{eqnarray}
and
\begin{eqnarray}\label{8}
p'=-\left(\rho c^{2} + p\right)\Phi'
\end{eqnarray}
Here the prime $'$ denotes the derivative with respect to $r$.\\
Now, if we review the vacuum solutions (including the Schwarzschild solution []) corresponding to the static, spherically symmetric metric in the framework of GR as well as MTGs then we can easily observe that the function $f(r)$ should be evaluated in terms of mass function $m(r)$. So, according to the ref.~\cite{kim2021neutron}, we have replaced $f(r)$ in terms of $m(r)$ as
\begin{eqnarray}\label{9}
f=\left[r\omega\left(r^{3}\omega + \frac{4G_{N}}{c^{2}}m\right)\right]^{1/2} - r^{2}\omega.
\end{eqnarray}
When $m(r)=M$ then $f$ approaches as $f\equiv \frac{2 G_{N}}{c^{2}}\frac{M}{r} - \frac{2 G_{N}^{2}}{c^{4}\omega}\frac{M}{r^{4}}+ ...$ corresponding to the limit $\omega \rightarrow \infty$. Where the first term is the same with the Schwarzschild solution \cite{schwarzschild1916gravitationsfeld} and the remaining terms are the HL corrections depending upon the parameter $\omega$.\\

Now, utilizing the Eqs.~(\ref{6})-(\ref{9}), we can established the following two first order non-linear differetial equations \cite{kim2021neutron}
\begin{eqnarray}\label{10}
\frac{dp}{dr}=\left(\rho c^{2} + p \right)\frac{r \omega \left[\left( 1+ \Tilde{\rho}\right) - \sqrt{1 + 4\Tilde{\rho}} - \Tilde{p}\right]}{\sqrt{1 + 4\Tilde{\rho}}\left[1 + r^{2}\omega\left(1 - \sqrt{1 + 4\Tilde{\rho}}\right)\right]},
\end{eqnarray}
\begin{eqnarray}\label{11}
\frac{dm}{dr}=4\pi r^{2}\rho,
\end{eqnarray}
where $\Tilde{\rho}=\frac{G_{N}}{c^{2}\omega}\frac{m}{r^{3}}$ and $\Tilde{p}=\frac{4\pi G_{N}}{c^{4}\omega}p$. The Eq.~(\ref{10}) known as the TOV equation, or hydrostatic equilibrium equation.\\
Now, using the relation $(1 + \sigma)^{1/2}\approx 1 + \frac{\sigma}{2} + ...$ for $\sigma<<1$ with take the leading term and the limit $\frac{1}{\omega}\rightarrow 0$, we can obtain usual TOV equation in GR as 
\begin{eqnarray}\label{12}
\frac{dp}{dr}\approx \frac{G_{N} m}{r^{2}c^{2}}\left(\rho c^{2} + p\right)\frac{\left(1 + \frac{4 \pi r^{3}}{m c^{2}}p\right)}{\left(1 - \frac{2 G_{N}m}{r c^{2}}\right)}.
\end{eqnarray}

The TOV equation can be described as the behavior of pressure $p(r)$ for a spherically symmetric object againts the radial coordinate $r$ when $p(r)$ is a function of its enclosed mass $m(r)$ and energy density $\rho(r)$. This equation firstly introduced in the res.~\cite{tolman1939static,oppenheimer1939massive} utilizing Einstein Field Equations and the Energy Momentum tensor. Again the TOV equation for an electrically charged object firstly discovered by Bekenstein \cite{bekenstein1971hydrostatic}. Now, an important fact that the interior structure for a static spherically symmetric stellar object made by one types of fluid can be fully described through the TOV equation. However, in the ref.~\cite{jimenez2022radial}, authors have explored features of quark stars with admixed two fluids through a couple of TOV equations. In the present model, our main motivation is to explore the features including mass-radius relation of proposed dark energy star by the modified TOV equation (\ref{10}) in HL gravity. Without loss of generality, we will analyze the rest portion of our current discussion by fixing the constant values as $G_{N}=c=1$.

\section{Equation of State}\label{sec3}
Till now, it is not exactly clear that the millisecond pulsars, observed by optical spectroscopic and photometric measurements are hadronic, strange/quark stars or hybrid stars. But theoretically possible existence of them are predicted by different models on strange stars and neutron(hybrid) stars \cite{annala2020evidence,demorest2010two,miller2019psr,cromartie2020relativistic,antoniadis2013massive,raaijmakers2019nicer,riley2019nicer}.
It is well known that as a possible alternative description of DE like phantom and quintessence fields, the chaplygin gas (CG) is the best DE candidate. The EoS obeying by CG takes the form $p=-\frac{B}{\rho}$ with $B$ being a positive constant in $km^{-4}$ units. Subsequently, a generalized chaplygin gas (GCG) has been introduced by several scientists, satisfying the EoS $p=-\frac{B}{\rho^{\alpha}}$ with an extra parameter $\alpha$ satisfying $0 \leq \alpha \leq 1 $. Although a lot literature \cite{bento2002generalized,cunha2004cosmological,gorini2008gauge,piattella2010extreme,reis2003entropy,salahedin2022new,von2023one,xu2012revisiting} have been explored in the context of the FLRW universe by introducing more generalized version of CG EoS, in our present study, the DE candidate is described by more generalized EoS (also known as modified chaplygin gas (MCG)) \cite{debnath2004role,zhai2006viscous,thakur2009modified} in the form 
\begin{eqnarray}\label{13}
p=A\rho - \frac{B}{\rho^{\alpha}} ~~~~\text{with}~~~~0 \leq \alpha \leq 1
\end{eqnarray}
Here the additional linear term $A\rho$ represents a barotropic fluid for positive $A$. For $\alpha =-1$ and $A=1 + B$, the standard cosmological constant type dark energy EOS can be described. Recently, using MCG and also GCG, several viable models \cite{bhar2020charged,malaver2022analytical,das2023dark,saleem2021charged,pretel2023radial,sepulveda2024radial} regarding physical characteristics of compact objects have been successfully explored. 

\section{Numerical Results}\label{sec4}

In order to obtain the interior solutions depicting the hydrostatic equilibrium position of our proposed dark energy stars, we need to solve the couple of differential equations (\ref{10}) and (\ref{11}). To close the coupled system of differential equations (\ref{10}) and (\ref{11}), we use an EoS (\ref{13}). Here we will integrate numerically by 4th-order Runge-Kutta method from the center towards the surface of the star with the initial conditions \cite{tangphati2022quark,tangphati2024anisotropic,li2023anisotropic}
\begin{eqnarray}
\rho(r=0)=\rho_{c}~~~~\text{and}~~~~m(r=0)=0,
\end{eqnarray}
where $\rho_{c}$ is the central energy density. The total mass is defined by $M\equiv m(r=R)$, $R$ is the radius of the star, determined through the condition $p(R)=0$ i.e., the fluid pressure vanishes at the surface. Here actually we will find three functions like $m$, $\rho$ and $p$ in terms of radial coordinate $r$. Now, due to high magnetic fields, extreme temperature and rotation properties, the internal structure of a compact relativistic object may be changed. However, a common fact of such objects that the observational mass-radius measurements for any type of EoS is used to categorize and comparison between different gravitation theories. In that perspective, we will emphasize the measurements of mass and radius for our proposed dark energy stars in HL gravity and also compare the results with observational measurements. Moreover, we will focus on the stability of our considered stellar system through various discussion. We begin our discussion corresponds three types of isotropic dark energy stars, classified through three different types of chaplygin gas EoS parameter like $\alpha=1$, $\alpha=0.9$ and $\alpha=0.8$. Without loss of generality, we adopt the values of constants $A = 0.45$ and $B = 0.16\times 10^{-6} km^{-4}$. It is clear that the pressure and energy density cannot be zero at the same time. So, for vanishing pressure at the surface 0f star, the non-vanishing energy density ia obtaines as $\left(B/A\right)^{1/(1 + \alpha)}$. It remarkable that the parameter $\omega$ is the only key parameter of our whole discussion by choosing its values as $\omega = 0.3, 0.4, 0.5, 0.6, 0.7, 0.8, 0.9$ in $10^{-1} km^{-2}$ units and $\omega \rightarrow \infty$ for GR. In the following subsections, we have reported some necessary features of compact stars in the perspective of our obtained stellar solutions:

\subsection{Pressure and Density}

\begin{figure}[!h]
  \centering
  \includegraphics[width=0.3\textwidth]{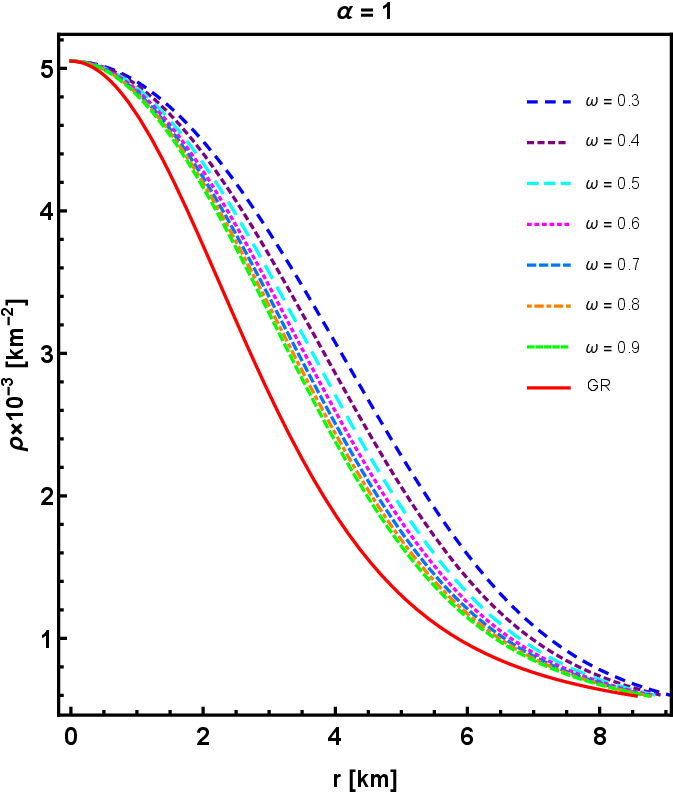}
  \hspace{2mm}
  \includegraphics[width=0.31\textwidth]{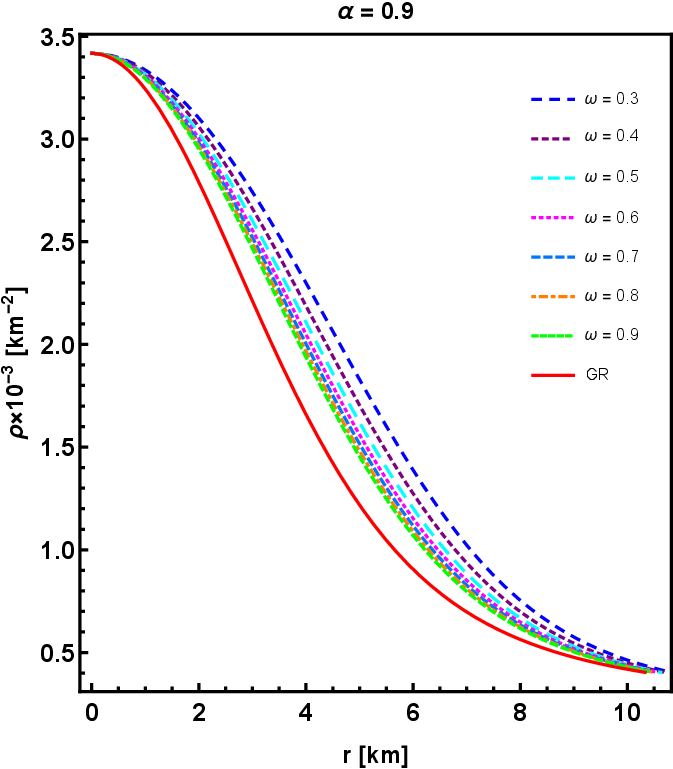}
  \hspace{2mm}
  \includegraphics[width=0.31\textwidth]{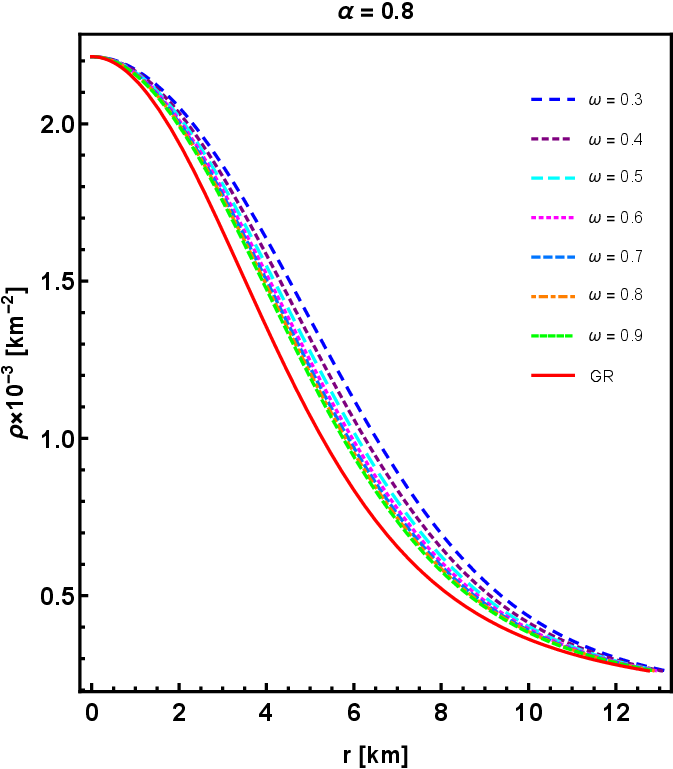}
  \caption{We display the behavior of energy density $\rho$ as a function of radial coordinate $r$ for three different variations of $\alpha$.} \label{fig1}
\end{figure}

\begin{figure}[thbp]
  \centering
  \includegraphics[width=0.31\textwidth]{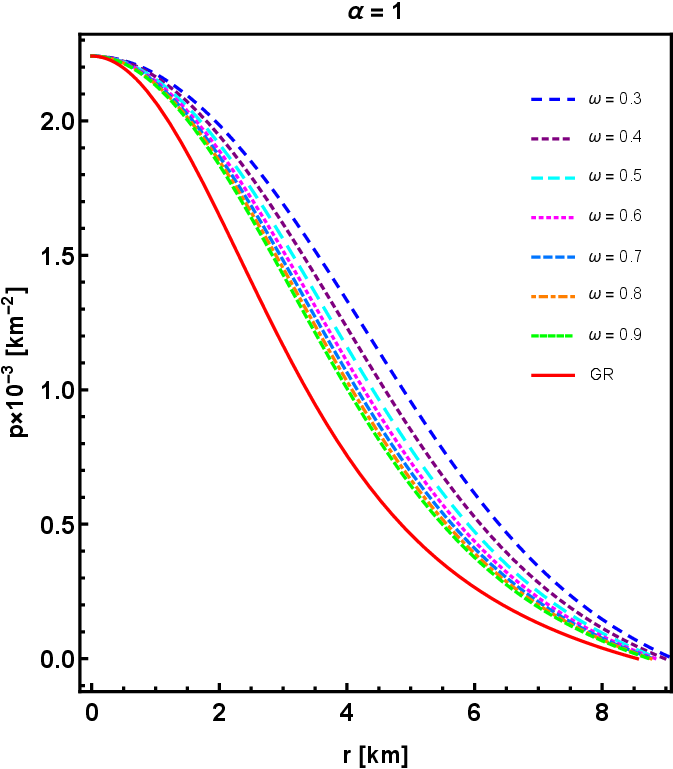}
  \hspace{2mm}
  \includegraphics[width=0.31\textwidth]{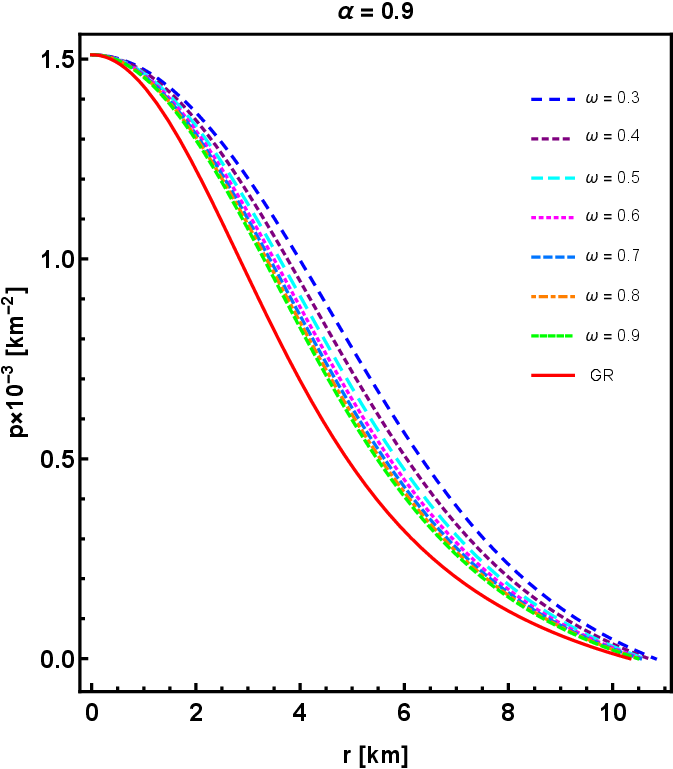}
  \hspace{2mm}
  \includegraphics[width=0.31\textwidth]{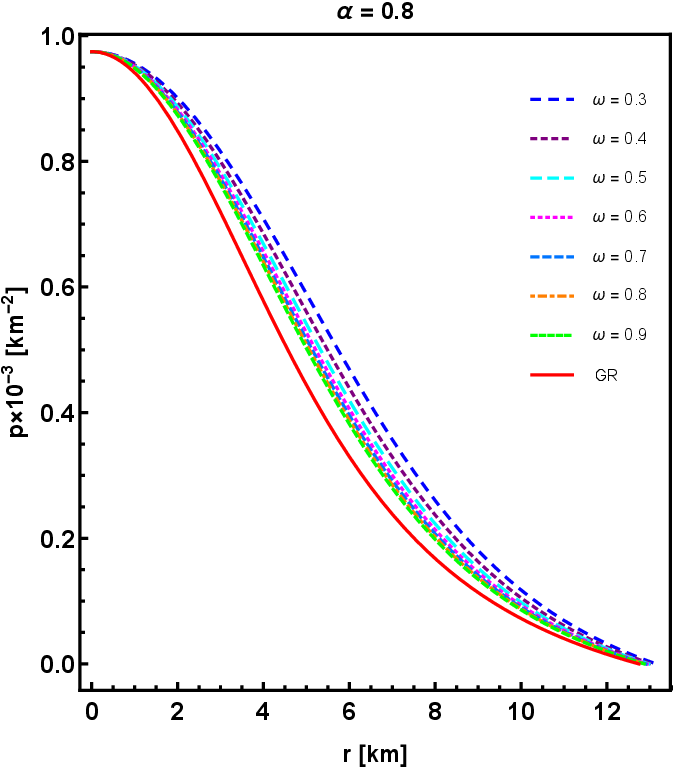}
  \caption{We display the behavior of pressure $p$ as a function of radial coordinate $r$ for three different variations of $\alpha$.} \label{fig2}
\end{figure}

Figs.~\ref{fig1} and \ref{fig2} illustrate the respective behavior of energy density/mass density $\rho$ and pressure $p$ as a function of radial coordinate $r$ corresponding to seven different values of the free parameter $\omega$. Also, in each panels of two Figs, we have shown the same behavior of $\rho$ and $p$ when $\omega\rightarrow \infty$ i.e., GR, depicted by \textit{red} solid line. One can easily observe that energy density and pressure are gradually decreases from the center towards the surface of stellar structure with both affected by $\omega$. Again the radius increases and central density decreases corresponding to less positive values of $\alpha$. Moreover, when $\omega$ increases the curve lines are deviate gradually towards the plot of GR and consequently, it must coincide with GR when $\omega \rightarrow \infty$. The same behavior also present for pressure profile which vanishes at surface of the star. Thus physically well-behaved nature for energy density and pressure for a compact stellar object are maintain in our considered dark energy stellar structure.

%%%%%%%%%%%%%%%%%%%%%%%%%%%%%%%%%%%%%%%%%
%%%%%%%%%%%%%%%%%%%%%%%%%%%%%%%%%%%%%%%%%
%%%%%%%%%%%%%%%%%%%%%%%%%%%%%%%%%%%%%%%

\subsection{Mass-Radius Relation}

\begin{figure}[thbp]
  \centering
  \includegraphics[width=0.31\textwidth]{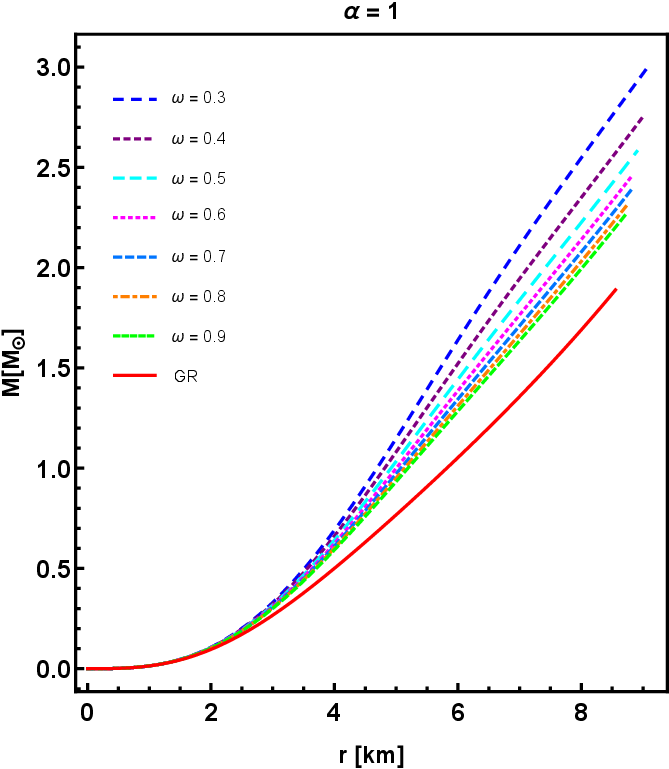}
  \hspace{2mm}
  \includegraphics[width=0.31\textwidth]{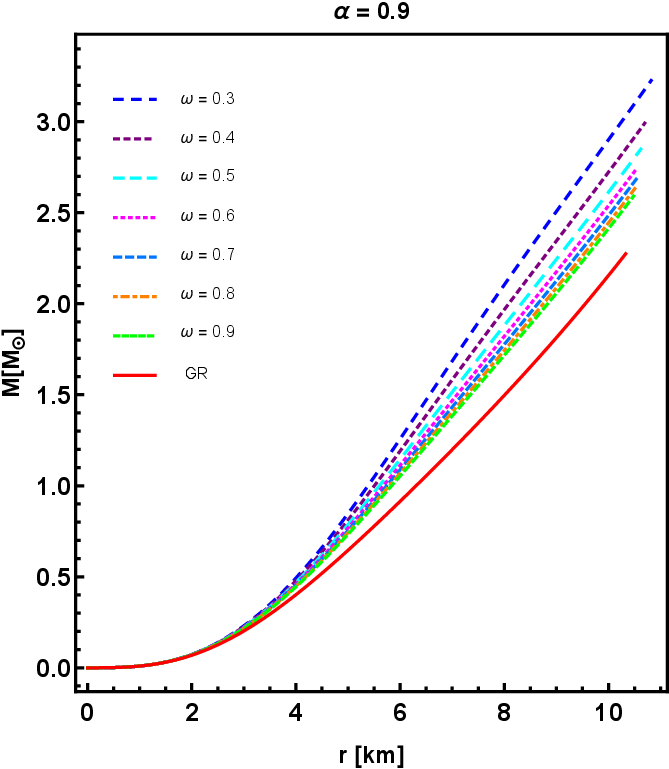}
  \hspace{2mm}
  \includegraphics[width=0.31\textwidth]{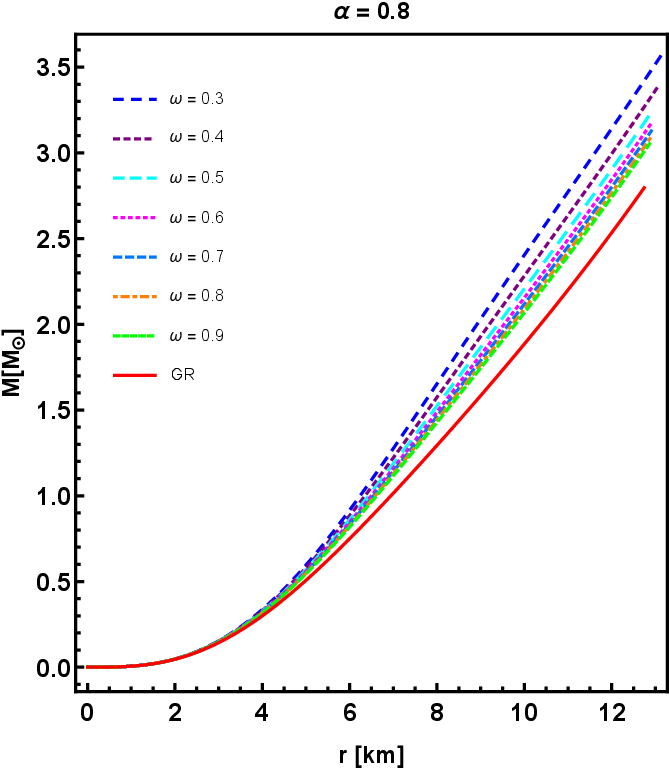}
  \caption{We display the behavior of mass $m(r)$ as a function of radial coordinate $r$ for three different variations of $\alpha$.} \label{fig3}
\end{figure}

\begin{figure}[thbp]
  \centering
  \includegraphics[width=0.47\textwidth]{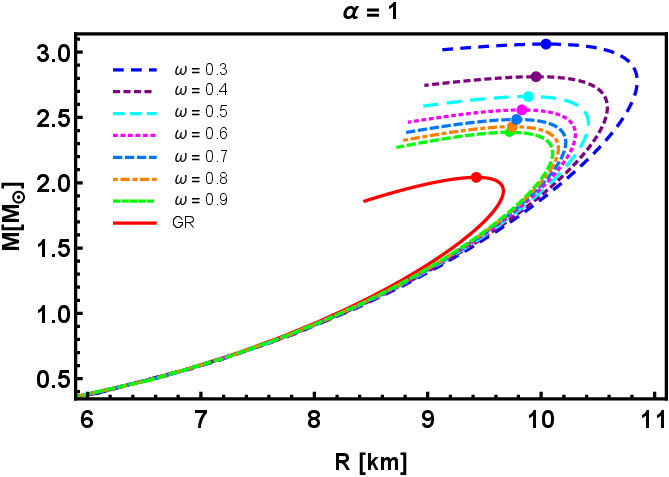}
  \hspace{2mm}
  \includegraphics[width=0.47\textwidth]{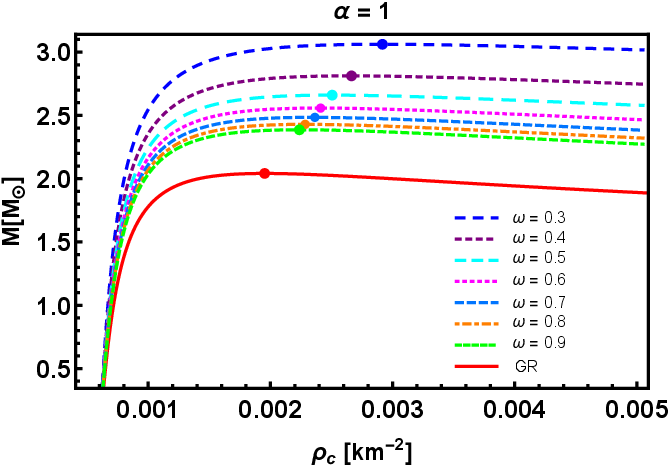}\\
   \vspace{2cm}
  \includegraphics[width=0.47\textwidth]{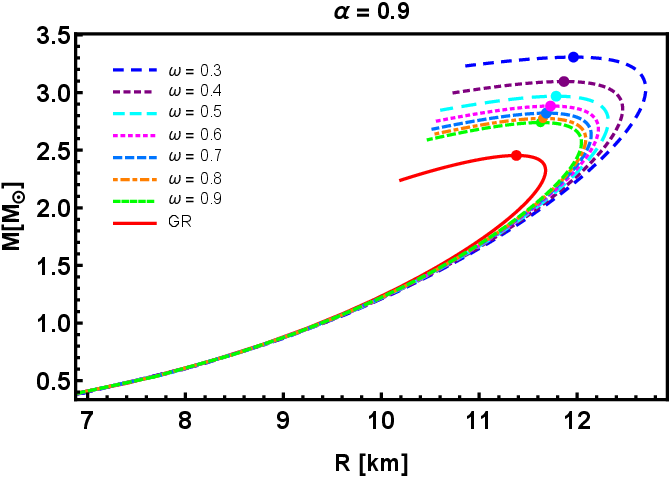}
   \hspace{2mm}
  \includegraphics[width=0.47\textwidth]{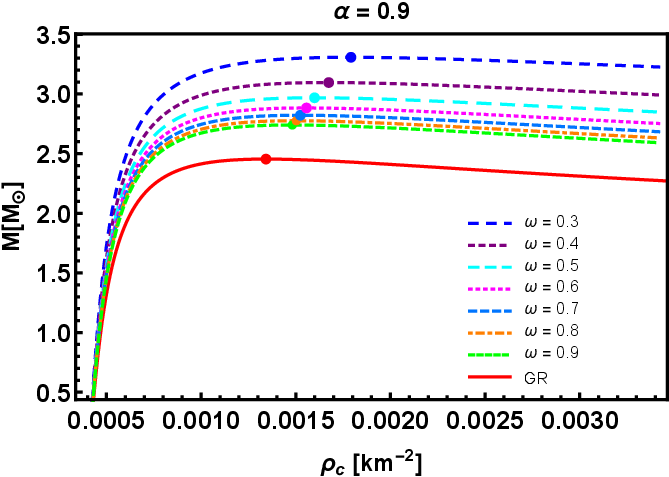}\\
   \vspace{1cm}
  \includegraphics[width=0.47\textwidth]{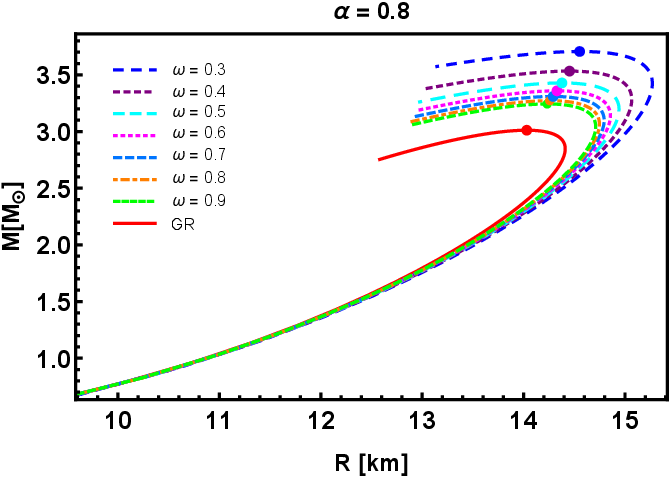}
   \hspace{2mm}
  \includegraphics[width=0.47\textwidth]{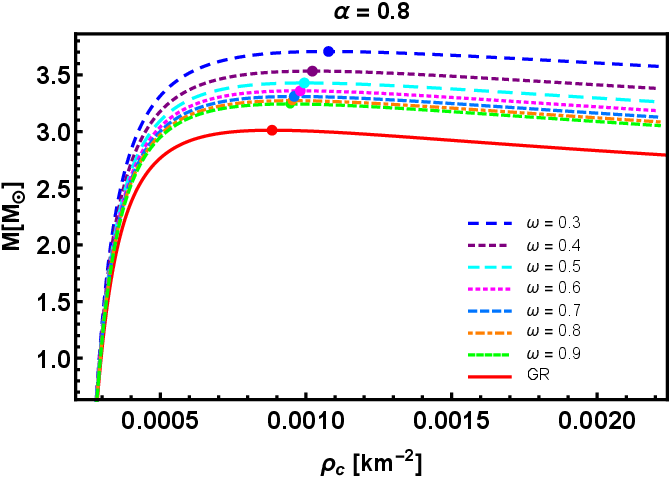}
  \caption{We display the behavior of mass ($M_{\odot}$)-radius ($R$) relation in the left panel of each row and mass ($M_{\odot}$)-central density ($\rho_{c}$)  in the right panel in each row for three different variations of $\alpha$.} \label{fig4}
\end{figure}

The investigation of mass-radius relation for analyzing a compact star model is most important because one can easily compared measured mass and radius with that of from various observed values. varying the central density $\rho_{c}$, we obtain the mass function $m(r)$, the maximum mass $M$ vs total radius $R$ relation and also compactness $M/R$ vs maximum mass $M$ relation. Note that we have measured the mass in solar mass units $M_{\odot}$, radius in kilometers $km$, whereas pressure and energy density in $km^{-2}$ units. In the Fig.~\ref{fig3}, we have actually shown the nature of mass while radius of the star increases in panels corresponds to $\alpha = 1,~ 0.9, ~0.8$. It is cleared that like a physically well-behaved compact stellar configuration, mass increases gradually from center towards the surface boundary for our considered dark energy stars corresponding to each variation of $\omega$. Next through the Fig.~\ref{fig4} we have illustrated the diagrams like maximum mass M versus total radius R in the left panel of each row and also maximum mass M versus corresponding central density $\rho_{c}$  in the right panel of each row for three variation of $\alpha = 1 (\text{upper row}),~ 0.9 (\text{middle row}) ~\text{and} 0.8 (\text{lower row})$. In each plot of the Fig.~\ref{fig4} corresponding to different values of $\omega$, we have marked the point $(M,R)$ by a big dot with the same color as the colored of the plot, which indicates the maximum mass M and corresponding predicted total radius of the star. For completeness, we have reported the maximum masses and corresponding predicted radii and also predicted central densities for each of seven variation of the model parameter $\omega$ with standard geometrized units in three tabular form like Table~\ref{table1} when $\alpha=1$, Table~\ref{table2} when $\alpha=0.9$ and Table~\ref{table3} when $\alpha=0.8$, obtained from the numerical calculation. By observing each plot and the numerical values, listed in the tables, one can easily find out the contribution of $\alpha$ as well as $\omega$ to estimate the maximum mass. It is remark that when $\alpha$ decreases then the maximum masses and corresponding radii are increases gradually for each variation of $\omega$. Again,   when $\omega$ increases then the values of M and R are both decrease corresponding to variation of $\alpha$. The same nature is also present for mass M versus central density $\rho_{c}$ in our current model. Now, if we focus on the deviation in HL from GR, then we have when $\omega$ increase to $\infty$ then all are coincide with that of GR. That means the theoretical aspect between HL and GR are also successfully presented in mass-radius and mass-central density relations for our considered dark energy stars model. \\
The compactness for compact relativistic objects/stars is an interesting feature, defined as $C\equiv M/R$. For our present model, we have demonstrated the plots $M-M/R$ in three panels correspond to $\alpha =1,~0.9,~0.8$ through the Fig.~\ref{fig4b}. Here we can also see that for different values of $\alpha$ compactness is increase when $\alpha$ decrease as well as $M/R$ decreases while $\omega$ increases. In the Tables~\ref{table1},~\ref{table2},~\ref{table3}, we have listed numerical values of the maximum compactness values corresponds to the maximum radius with each variation of $\omega$. By observing these values, it is notable that compactness is also consistent with $M-R$ for our current model. Further, compactness satisfies physically well-behaved Buchdahl condition \cite{buchdahl1959general} for compact object because $M/R<4/9$ hold good in our model.\\
For a Schwarzschild stellar structure, the surface redshift $Z{sur}$ can be defined as \cite{glendenning2012compact}
\begin{eqnarray}
Z_{sur}=\left(1-\frac{2M}{R}\right)^{-1/2}-1.
\end{eqnarray}
According to Barraco and Hamity \cite{barraco2002maximum}, the value of $Z_{sur}$ for an isotropic star in absence of cosmological constant must satisfy the viability condition $Z_{sur}<2$ whereas for anisotropic with cosmological constant it satisfy $Z_{sur}<5$ \cite{bohmer2006bounds}. Later Ivanov \cite{ivanov2002maximum} proposed the maximum acceptable value of surface redshift as 5.211. By the Fig.~\ref{fig4b}, the plot $Z_{sur}-M$ have been demonstrated in three panels corresponds to three different values of $\alpha$ like $1,~0.9,~0.8$. Also, in the Tables~\ref{table1},~\ref{table2},~\ref{table3}, we have listed the numerical maximum values of $Z_{sur}$ for each variation of $\omega$. It is remark that our dark energy star model satisfy the viability condition by surface redshift as $Z_{sur}<1$ for each case of $\alpha= 1,~0.9,~0.8$. \\
Moreover, it is predicted that a compact object with electrically charged or with anisotropic matter can achieve more masses as compared with neutral isotropic object \cite{panotopoulos2019relativistic,lopes2019anisotropic}. Now, it is confirmed that our results are significantly affect by specific EoS (\ref{13}) in HL gravity. In comparing our obtained results with recent observational data, we will use observational constraints of some massive pulsar like NSs. For example,  $PSR J0952-0607$ with mass $2.35 \pm 0.17$ $M_{\odot}$ \cite{romani2022psr}, $PSR J0348+0432$ with mass $2.01 \pm 0.04$ $M_{\odot}$ \cite{antoniadis2013massive}, pulsars $J1614-2230$ with mass $1.928 \pm 0.017$ $M_{\odot}$ \cite{demorest2010shapiro}, $PSR J0751+1807$ with mass $2.1\pm 0.2$ $M_{\odot}$  \cite{nice20052}.
Also, $PSR J0740+6620$ with mass $2.08 \pm 0.07$ $M_{\odot}$ from NICER and XMM-Newton Data \cite{miller2021radius} and $2.14 ^{+0.2}_{-0.17}$ $M_{\odot}$ measured in the ref.~\cite{cromartie2020relativistic}.  Now, for our model, the maximum mass $M_{max}\in [2.3,3]$  $M_{\odot}$ and gradually decreases when $\omega$ increases, details are reported in the Table~\ref{table1}.\\
So, in particular for a comparatively large value of $\omega$, we can predict the existence of massive pulsars like $PSR J0952-0607$, $PSR J0740+6620$, and $PSR J0751+1807$ in the case of $\alpha = 1$ in HL gravity. In similar way for other two cases like $\alpha=0.9$ and $\alpha = 0.8$, we can predict the existence of several massive pulsars by adjusting the parameter $\omega$ in the context of HL gravity. Therefore, in context of HL gravity, our proposed dark energy stars model physically reasonable.\\
The compactness of a compact star indicates the strongness of gravitational field. Again, a compact object collapsing into a black hole when the total radius becomes less than its Schwarzschild radius $R_{Sch}$. In our present study, we will find the values of $R_{Sch}$ in km units by the condition $f(R_{Sch})=1$ \cite{eslam2019white,hendi2017neutron}. In Tables~\ref{table1},~\ref{table2}, and~\ref{table3}, we have reported the values of $R_{Sch}$ for each variatiation of $\omega$. One can easily find that $R_{Sch}<R$ and which demands that our considered stellar system does not collapse into a black hole in HL gravity.

%%%%%%%%%%%%%%%%%%%%%%%%%%%%%%%%%%%%%%%%%
%%%%%%%%%%%%%%%%%%%%%%%%%%%%%%%%%%%%%%%
%%%%%%%%%%%%%%%%%%%%%%%%%%%%%%%%%%%%%%%%%
%%%%%%%%%%%%%%%%%%%%%%%%%%%%%%%%%%%%%%%%%
%%%%%%%%%%%%%%%%%%%%%%%%%%%%%%%%%%%%%%%%%
\begin{figure}[thbp]
  \centering
  \includegraphics[width=0.31\textwidth]{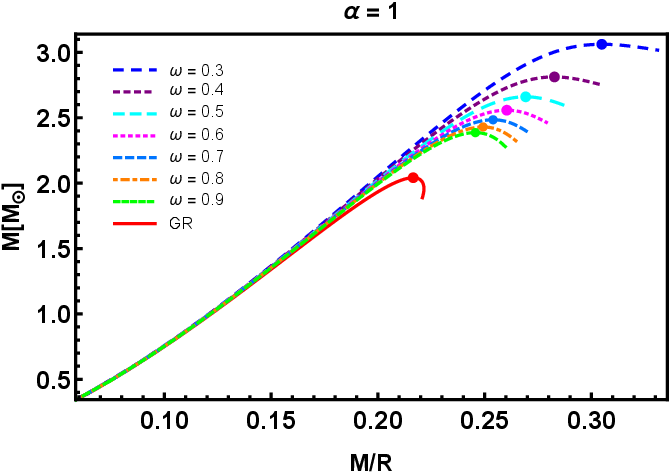}
  \hspace{2mm}
  \includegraphics[width=0.31\textwidth]{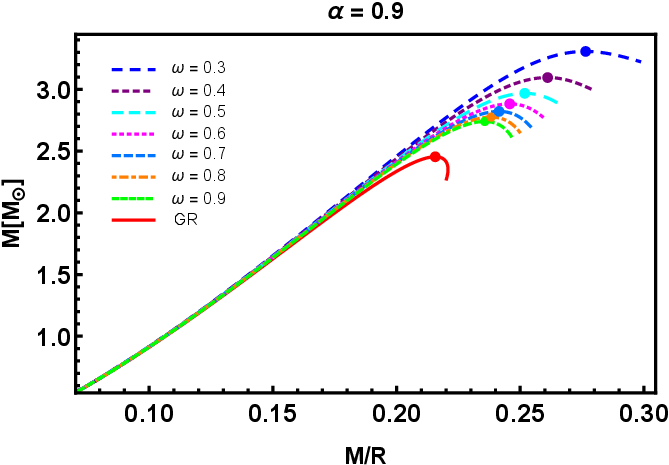}
  \hspace{2mm}
  \includegraphics[width=0.31\textwidth]{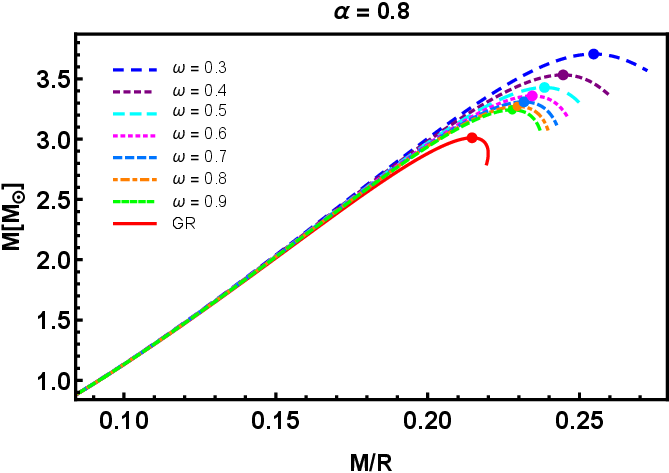}
  \caption{Represents the behavior of Mass M$[M_{\odot}]$  vs compactness $M/R$ for three different variations of $\alpha$.} \label{fig4a}
\end{figure}
%%%%%%%%%%%%%%%%%%%%%%%%%%%%%%%%%%%%%%%
%%%%%%%%%%%%%%%%%%%%%%%%%%%%%%%%%%%%%%%%%
\begin{figure}[thbp]
  \centering
  \includegraphics[width=0.31\textwidth]{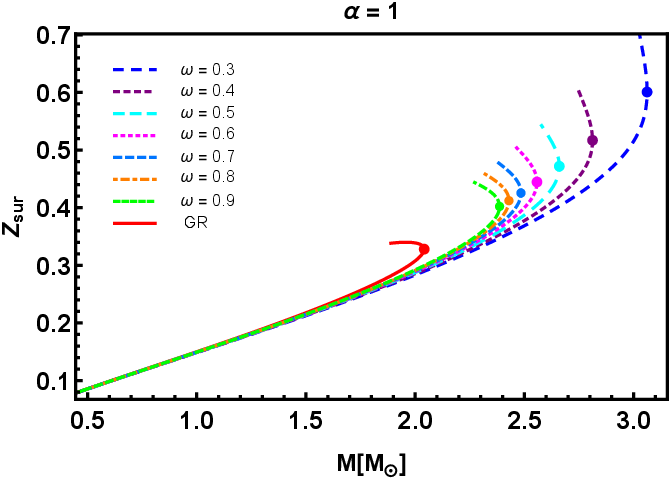}
  \hspace{2mm}
  \includegraphics[width=0.31\textwidth]{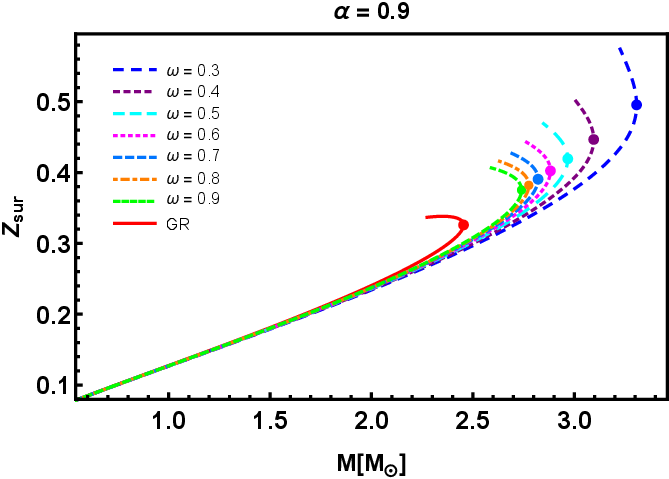}
  \hspace{2mm}
  \includegraphics[width=0.31\textwidth]{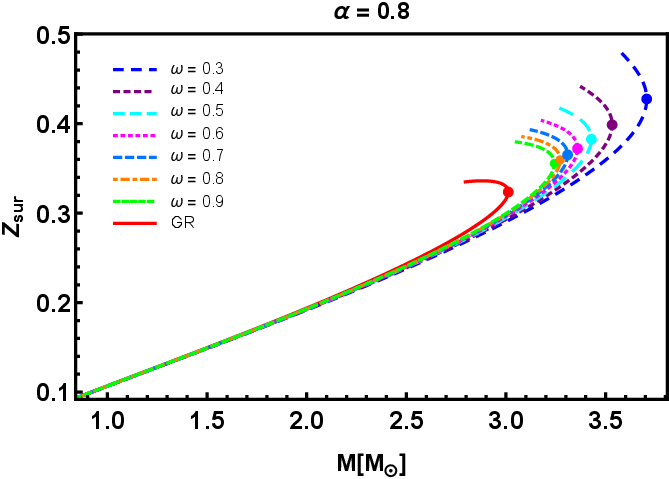}
  \caption{Represents the behavior of surface redshift $Z_{sur}$ vs Mass M$[M_{\odot}]$ for three different variations of $\alpha$.} \label{fig4b}
\end{figure}
%%%%%%%%%%%%%%%%%%%%%%%%%%%%%%%%%%%%%%%
%%%%%%%%%%%%%%%%%%%%
\begin{table}
  \centering
  \begin{tabular}{ |p{2cm}|p{2.3cm}|p{2.3cm}|p{2.8cm}|p{2cm}|p{2cm}|p{2cm}| }
\hline
 \multicolumn{7}{|c|}{\textbf{\textit{\large{$\alpha=1$}} }} \\[1.3mm]
 \hline
  \centering $\boldsymbol{\omega [10^{-1}km^{-2}]}$ & \centering $\boldsymbol{M_{max}[M_{\odot}]}$ & \centering $\boldsymbol{R_{predicted}[km]}$ & \centering $\boldsymbol{(\rho_{c})_{predicted}[10^{14}\frac{g}{cm^{3}}]}$ & \centering $\boldsymbol{(M/R)_{max}}$ & \centering $\boldsymbol{(Z_{sur})_{max}}$  & $\boldsymbol{R_{Sch}[km]}$\\[0.8mm]
 \hline
  \centering 0.3   &   \centering 3.06178   &  \centering 10.0427  &  \centering 1.566856  &  \centering 0.304876 &\centering 0.600771 & ~~~ 6.46446 \\[1.1mm]
  \hline
  \centering 0.4  & \centering 2.8121   & \centering 9.95469 &  \centering 1.431074 &  \centering 0.282705 & \centering 0.51691 &~~~ 6.33077 \\[1.1mm]
  \hline
  \centering 0.5   & \centering 2.65997   & \centering 9.88747 & \centering 1.346172 &  \centering 0.269225 & \centering 0.471942 &~~~ 6.25800\\[1.1mm]
  \hline
 \centering 0.6  &  \centering 2.55776   & \centering 9.82998 & \centering 1.295245 &  \centering  0.260392 & \centering 0.444556 & ~~~6.21276 \\[1.1mm]
 \hline
 \centering 0.7  &  \centering 2.48444   & \centering 9.78296 & \centering 1.269779 &  \centering  0.253956 & \centering 0.425538 & ~~~6.18214 \\[1.1mm]
 \hline
 \centering 0.8  &  \centering 2.42930   & \centering 9.74553 & \centering 1.227338 &  \centering 0.249274 & \centering 0.412163 & ~~~6.1601 \\[1.1mm]
 \hline
 \centering 0.9  &  \centering 2.38634   & \centering 9.71978 & \centering 1.201877 &  \centering  0.245691 & \centering 0.402181 & ~~~6.14352 \\[1.1mm]
 \hline
  \centering $\rightarrow\infty$ ($\equiv$ GR)  &  \centering 2.04146   & \centering 9.427885 & \centering 1.049095 &  \centering  0.216535 & \centering 0.328113 & ~~~$\approx$6.02925 \\[1.1mm]
 \hline
\end{tabular}
  \caption{We reported the maximum values of mass $M_{max}$ in solar mass unit $M_{\odot}$, corresponding predicted radius $R_{predicted}$ in kilometer unit, corresponding predicted central density $(\rho_{c})_{predicted}$, maximum compactness $(M/R)_{max}$, maximum surface redshift $(Z_{sur})_{max}$ and the Schwarzschild radius $R_{Sch}$ in kilometer unit for parameter $\alpha=1$ when $\omega\in[0.3,0.9]$ with $\delta\omega=0.1$ and $\omega\rightarrow \infty(\equiv$GR).}\label{table1}
\end{table}
%%%%%%%%%%%%%%%%%%%%%%%%%%%%%%%%%%%
%%%%%%%%%%%%%%%%%%%%
\begin{table}
  \centering
  \begin{tabular}{ |p{2cm}|p{2.3cm}|p{2.3cm}|p{2.8cm}|p{2cm}|p{2cm}|p{2cm}| }
\hline
 \multicolumn{7}{|c|}{\textbf{\textit{\large{$\alpha=0.9$}} }} \\[1.3mm]
 \hline
  \centering $\boldsymbol{\omega [10^{-1}km^{-2}]}$ & \centering $\boldsymbol{M_{max}[M_{\odot}]}$ & \centering $\boldsymbol{R_{predicted}[km]}$ & \centering $\boldsymbol{(\rho_{c})_{predicted}[10^{13}\frac{g}{cm^{3}}]}$ & \centering $\boldsymbol{(M/R)_{max}}$ & \centering $\boldsymbol{(Z_{sur})_{max}}$  & $\boldsymbol{R_{Sch}[km]}$\\[0.8mm]
 \hline
  \centering 0.3   &   \centering 3.30715   &  \centering 11.963  &  \centering 9.62404  &  \centering 0.275448 &\centering 0.495532 & ~~~ 7.56388 \\[1.1mm]
  \hline
  \centering 0.4  & \centering 3.09567   & \centering 11.8656 &  \centering 8.9924 &  \centering 0.261082 & \centering 0.446639 &~~~ 7.46922 \\[1.1mm]
  \hline
  \centering 0.5   & \centering 2.96792   & \centering 11.7861 & \centering 8.59045 &  \centering 0.251815 & \centering 0.419375 &~~~ 7.41725\\[1.1mm]
  \hline
 \centering 0.6  &  \centering 2.88248   & \centering 11.7283 & \centering 8.36077 &  \centering  0.24577 & \centering 0.40240 & ~~~7.38465 \\[1.1mm]
 \hline
 \centering 0.7  &  \centering 2.82135   & \centering 11.6859 & \centering 8.1885 &  \centering  0.241431 & \centering 0.390583 & ~~~7.36239 \\[1.1mm]
 \hline
 \centering 0.8  &  \centering 2.77546  & \centering 11.6507 & \centering 8.07361 &  \centering  0.238222 & \centering 0.382032 & ~~~7.34627 \\[1.1mm]
 \hline
 \centering 0.9  &  \centering 2.73975   & \centering 11.6263 & \centering 7.95876 &  \centering  0.2356521 & \centering 0.375298 & ~~~7.33408 \\[1.1mm]
 \hline
  \centering $\rightarrow\infty$ ($\equiv$ GR)  &  \centering 2.45394   & \centering 11.380 & \centering 7.21228 &  \centering  0.215637 & \centering 0.326015 & ~~~$\approx$7.24747 \\[1.1mm]
 \hline
\end{tabular}
  \caption{We reported the maximum values of mass $M_{max}$ in solar mass unit $M_{\odot}$, corresponding predicted radius $R_{predicted}$ in kilometer unit, corresponding predicted central density $(\rho_{c})_{predicted}$, maximum compactness $(M/R)_{max}$, maximum surface redshift $(Z_{sur})_{max}$ and the Schwarzschild radius $R_{Sch}$ in kilometer unit for parameter $\alpha=0.9$ when $\omega\in[0.3,0.9]$ with $\delta\omega=0.1$ and $\omega\rightarrow \infty(\equiv$GR).}\label{table2}
\end{table}
%%%%%%%%%%%%%%%%%%%%%%%%%%%%%%%%%%%
%%%%%%%%%%%%%%%%%%%%
\begin{table}
  \centering
  \begin{tabular}{ |p{2cm}|p{2.3cm}|p{2.3cm}|p{2.8cm}|p{2cm}|p{2cm}|p{2cm}| }
\hline
 \multicolumn{7}{|c|}{\textbf{\textit{\large{$\alpha=0.8$}} }} \\[1.3mm]
 \hline
  \centering $\boldsymbol{\omega [10^{-1}km^{-2}]}$ & \centering $\boldsymbol{M_{max}[M_{\odot}]}$ & \centering $\boldsymbol{R_{predicted}[km]}$ & \centering $\boldsymbol{(\rho_{c})_{predicted}[10^{13}\frac{g}{cm^{3}}]}$ & \centering $\boldsymbol{(M/R)_{max}}$ & \centering $\boldsymbol{(Z_{sur})_{max}}$  & $\boldsymbol{R_{Sch}[km]}$\\[0.8mm]
 \hline
  \centering 0.3   &   \centering 3.70619   &  \centering 14.5530  &  \centering 5.78792  &  \centering 0.254668 &\centering 0.427606 & ~~~ 9.11797 \\[1.1mm]
  \hline
  \centering 0.4  & \centering 3.53282   & \centering 14.4528 &  \centering 5.49034 &  \centering 0.244614 & \centering 0.39874 &~~~ 9.05309 \\[1.1mm]
  \hline
  \centering 0.5   & \centering 3.42859   & \centering 14.3782 & \centering 5.34156 &  \centering 0.238458 & \centering 0.382656 &~~~ 9.01698 \\[1.1mm]
  \hline
 \centering 0.6  &  \centering 3.35904  & \centering 14.3292 & \centering 5.26717 &  \centering  0.23442 & \centering 0.372104 & ~~~8.99405 \\[1.1mm]
 \hline
 \centering 0.7  &  \centering 3.30935   & \centering 14.2914 & \centering 5.15557 &  \centering  0.231726 & \centering 0.365198 & ~~~8.97826 \\[1.1mm]
 \hline
 \centering 0.8  &  \centering 3.27209   & \centering 14.2573 & \centering 5.11837 &  \centering  0.229502 & \centering 0.359575 & ~~~8.96677 \\[1.1mm]
 \hline
 \centering 0.9  &  \centering 3.24311  & \centering 14.2327 & \centering 5.08118 &  \centering  0.227864 & \centering 0.355476 & ~~~8.95802 \\[1.1mm]
 \hline
  \centering $\rightarrow\infty$ ($\equiv$ GR)  &  \centering 3.01146   & \centering 14.0313 & \centering 4.77185 &  \centering  0.2146251 & \centering 0.323661 & ~~~$\approx$8.89405 \\[1.1mm]
 \hline
\end{tabular}
  \caption{We reported the maximum values of mass $M_{max}$ in solar mass unit $M_{\odot}$, corresponding predicted radius $R_{predicted}$ in kilometer unit, corresponding predicted central density $(\rho_{c})_{predicted}$, maximum compactness $(M/R)_{max}$, maximum surface redshift $(Z_{sur})_{max}$ and the Schwarzschild radius $R_{Sch}$ in kilometer unit for parameter $\alpha=0.8$ when $\omega\in[0.3,0.9]$ with $\delta\omega=0.1$ and $\omega\rightarrow \infty(\equiv$GR).}\label{table3}
\end{table}
%%%%%%%%%%%%%%%%%%%%%%%%%%%%%%%%%%%

\subsection{Stability}
Along with mass-radius relation the most important issue for analyzing the compact stellar system is stability of the considered system. Below we investigate the stability of our proposed dark energy stellar structure in HL gravity in details.
\subsubsection{Via Sound Speed}
\begin{figure}[thbp]
  \centering
  \includegraphics[width=0.31\textwidth]{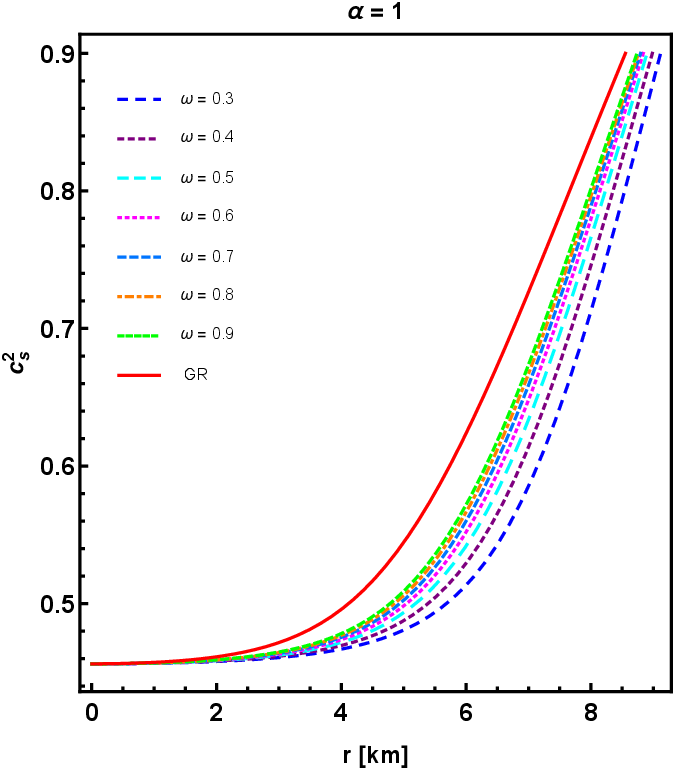}
  \hspace{2mm}
  \includegraphics[width=0.31\textwidth]{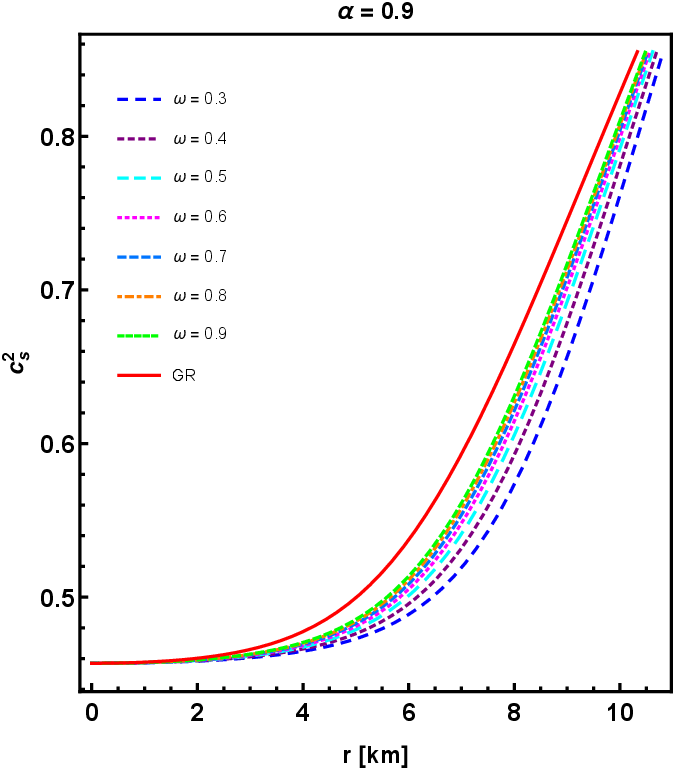}
  \hspace{2mm}
  \includegraphics[width=0.32\textwidth]{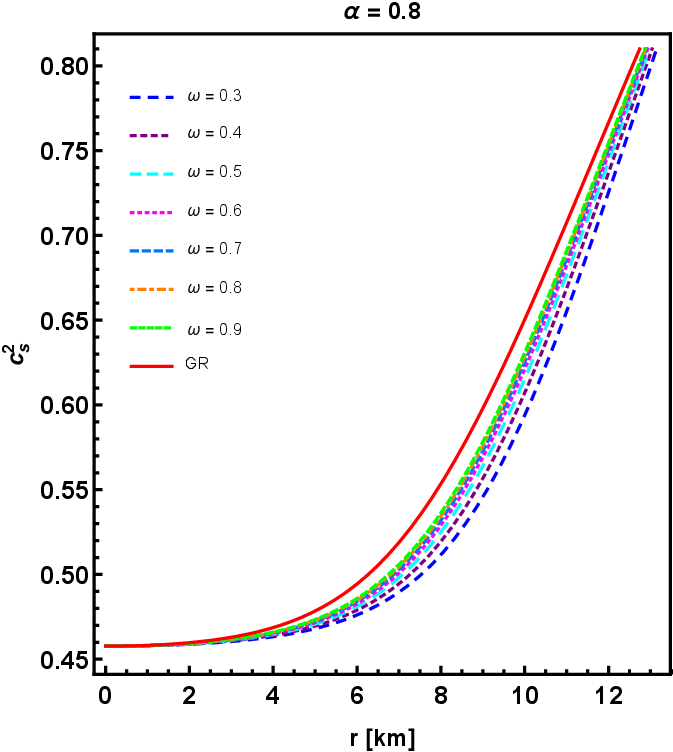}
  \caption{We display the behavior of sound speed $c_{s}^{2}$ vs radial coordinate $r$ for three different variations of $\alpha$.} \label{fig5}
\end{figure}

In the Fig.~\ref{fig5}, squared sound speed $c_{s}^{2}=\frac{dp}{d\rho}$ has been depicted as a function of radius through three panels for three different values of $\alpha$. We can easily verify that $c_{s}^{2}$ increases from center towards the surface of star for each different values of $\omega$ as well as GR. Also, in the whole interior region of stellar system, $c_{s}^{2}$ fulfill the causality condition i.e., $0 < c_{s}^{2} < 1$ \cite{abreu2007sound}. So, the physically stability behavior are presented in our proposed stellar system.  
%%%%%%%%%%%%%%%%%%%%%%%%%%%%%%%%%%%%%%%%%
%%%%%%%%%%%%%%%%%%%%%%%%%%%%%%%%%%%%%%%
\subsubsection{Dynamical Stability}
\begin{figure}[thbp]
  \centering
  \includegraphics[width=0.31\textwidth]{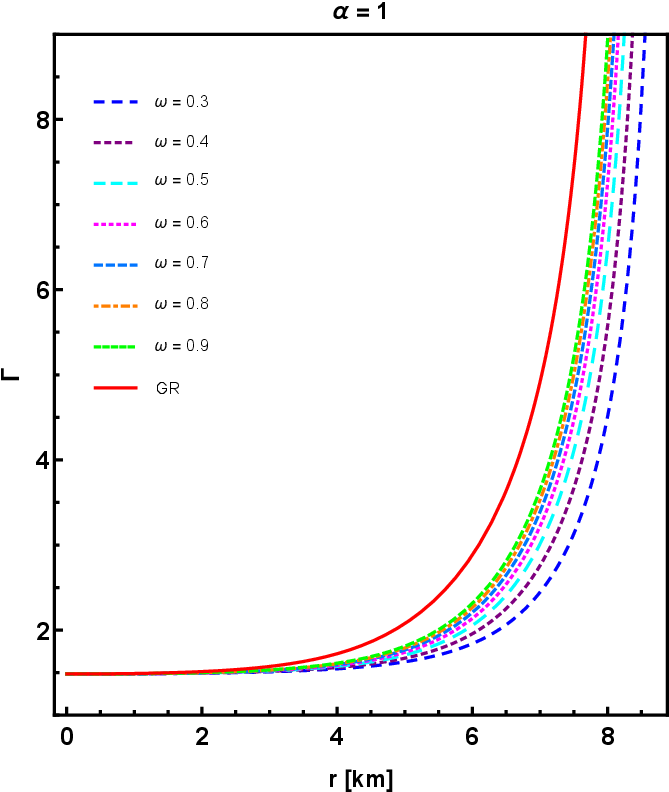}
  \hspace{2mm}
  \includegraphics[width=0.31\textwidth]{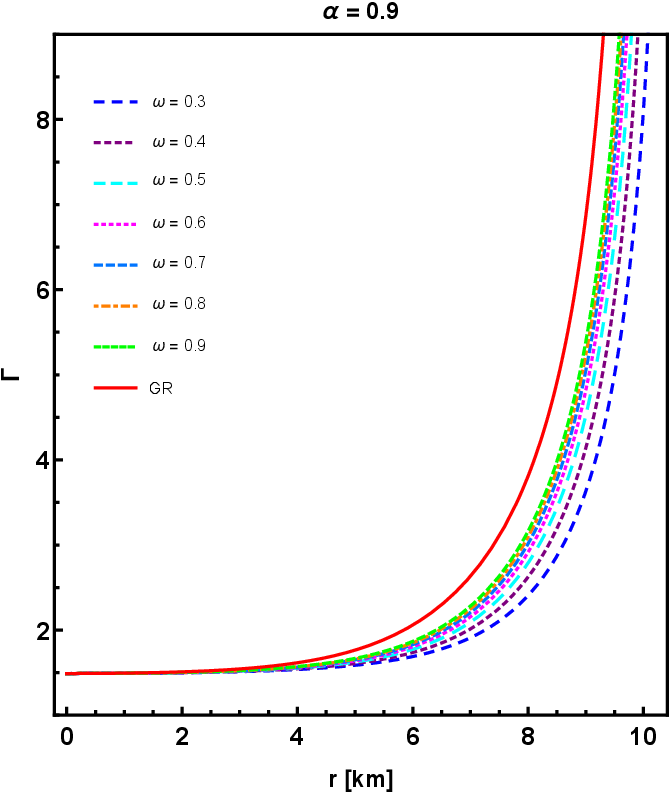}
  \hspace{2mm}
  \includegraphics[width=0.31\textwidth]{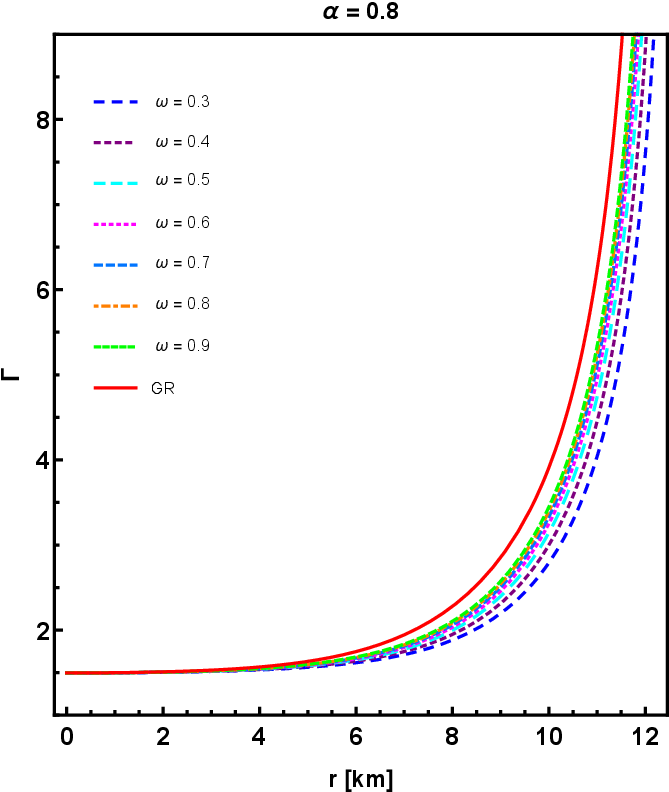}
  \caption{We display the behavior of adiabatic index $\Gamma$ vs radial coordinate $r$ for three different variations of $\alpha$.} \label{fig6}
\end{figure}

Now, we will investigate the dynamical stability of our considered stellar system by examining the adiabatic index $Gamma$. Chandrasekhar as a pioneer first introduced such dynamical stability criteria of a stellar system against infinitesimal radial perturbation \cite{chandrasekhar1965stability}. Adiabatic index defined as
\begin{eqnarray}\label{25}
\Gamma=\left(1 + \frac{\rho}{p}\right)\frac{dp}{d\rho},
\end{eqnarray} 
where $\frac{dp}{d\rho}$ is the  squared sound speed $c_{s}^{2}$. Remark that $\Gamma$ is a dimensionless quantity indicating the stiffness of the EoS. According to the refs.~\cite{glass1983stability,moustakidis2017stability}, a stellar system will be stable when at each point of whole interior region $\Gamma$ is greater than $4/3$ otherwise stellar configurations will be unstable against radial perturbation. Finally, through the Fig.~\ref{fig6}, we display the plot of $\Gamma$ as a function of $r$ in three panels for three different values of $\alpha$. As a result, we have $\Gamma>4/3\sim 1.33$ for each different values of the parameter $\omega$ and also for GR. So, our proposed model is stable against infinitesimal radial perturbation and increasing behavior of $\Gamma$ indicates the growth of pressure corresponding to increase values of energy density i.e., a stiffer EoS.

%%%%%%%%%%%%%%%%%%%%%%%%%%%%%%%%%%%%%%%%%
%%%%%%%%%%%%%%%%%%%%%%%%%%%%%%%%%%%%%%%

Further, the stability condition for a compact star including dark energy star, known as Harrison-Zeldovich-Novikov criterion \cite{harrison1965gravitation,zel1972relativistic}, is as follows
\begin{eqnarray}\label{26}
\frac{dM}{d\rho_{c}}>0~~~\rightarrow~~~\text{stable configuration}~~~~\text{and}~~~~\frac{dM}{d\rho_{c}}<0~~~\rightarrow~~~\text{unstable configuration}.
\end{eqnarray}
In particular, from the right panels of the Fig.~\ref{fig4}, we can easily observe that mass $M$ increases with the variation of central density $\rho_{c}$ until it reaches the maximum value $M_{max}$ and after that value mass becomes in decreasing nature with central density. Thus, $M_{max}$, indicated by big dot in each plot, represents the dividing point between stable and unstable configuration of our proposed dark energy stars model in HL gravity.

\subsection{Radial Oscillations}
\begin{figure}[thbp]
  \centering
  \includegraphics[width=0.47\textwidth]{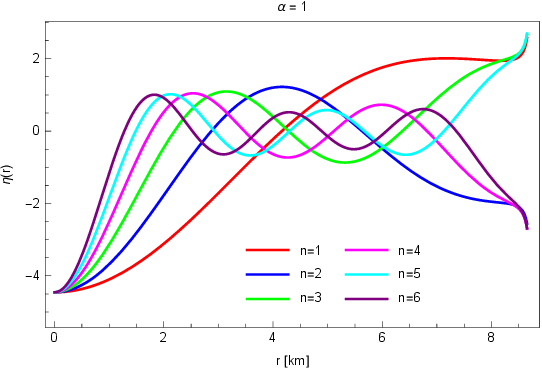}
  \hspace{2mm}
  \includegraphics[width=0.47\textwidth]{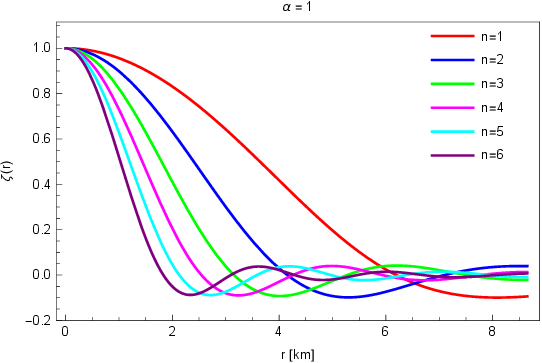}\\
   \vspace{0.5cm}
  \includegraphics[width=0.47\textwidth]{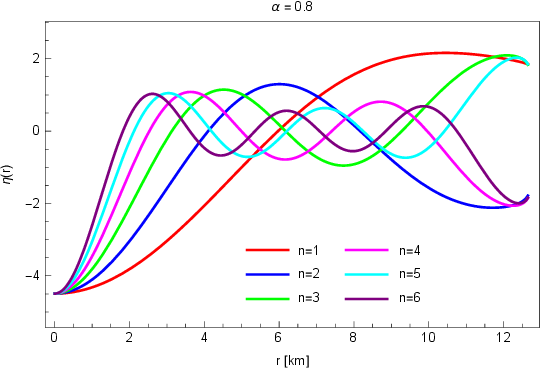}
   \hspace{2mm}
  \includegraphics[width=0.47\textwidth]{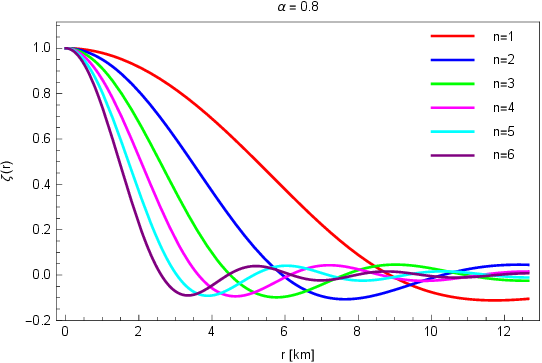}\\
   \vspace{0.7cm}
  \includegraphics[width=0.47\textwidth]{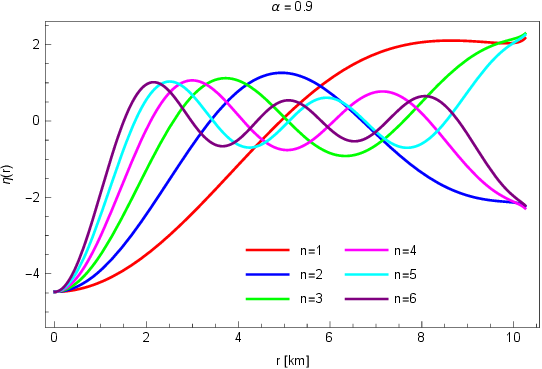}
   \hspace{2mm}
  \includegraphics[width=0.47\textwidth]{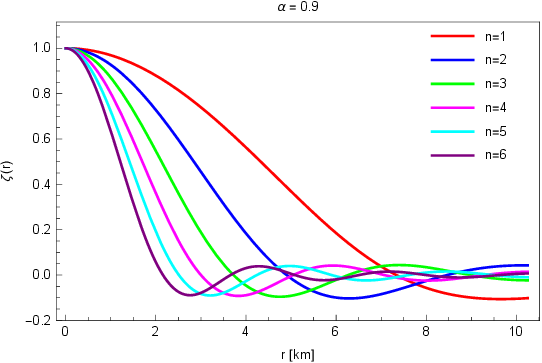}
  \caption{We display the behavior of eigenfunctions $\eta(r)$ in the left panels of each rows and $\zeta(r)$ in the right panels of each rows against the radial coordinate $r$ when $\omega=0.9 [10^{-1}km^{-2}]$ corresponding to three variations of $\alpha=1, 0.8, 0.9$.} \label{fig7}
\end{figure}

\begin{figure}[thbp]
  \centering
  \includegraphics[width=0.47\textwidth]{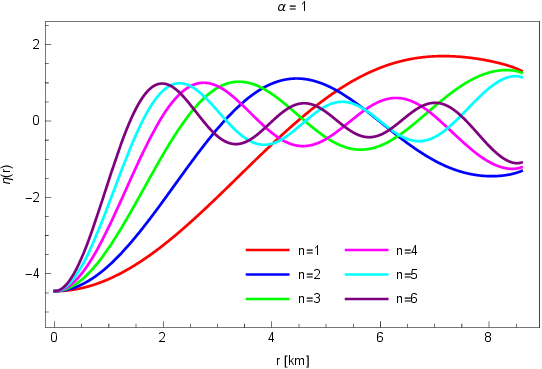}
  \hspace{2mm}
  \includegraphics[width=0.47\textwidth]{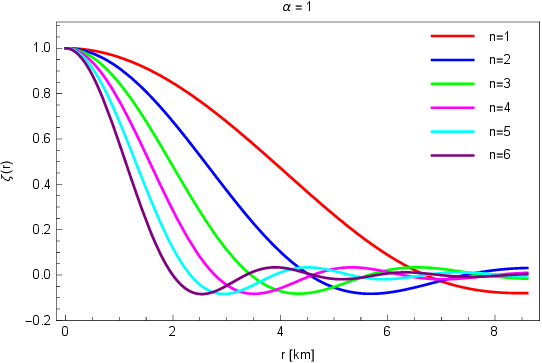}\\
   \vspace{0.5cm}
  \includegraphics[width=0.47\textwidth]{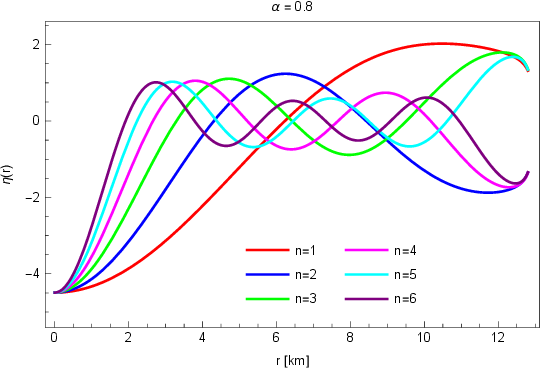}
   \hspace{2mm}
  \includegraphics[width=0.47\textwidth]{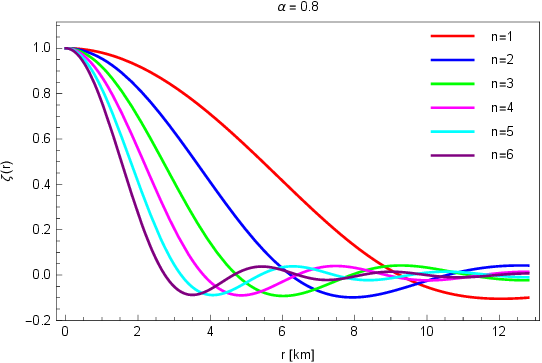}\\
   \vspace{0.7cm}
  \includegraphics[width=0.47\textwidth]{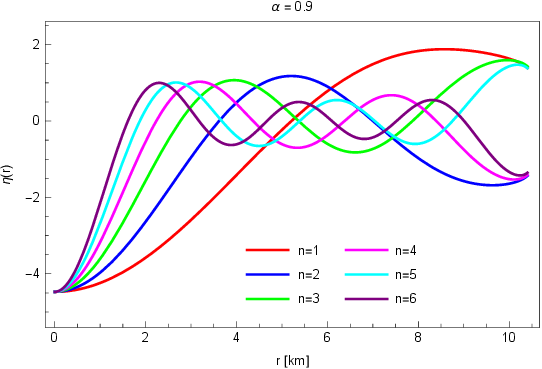}
   \hspace{2mm}
  \includegraphics[width=0.47\textwidth]{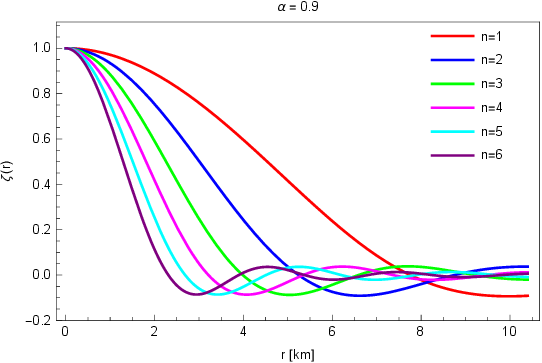}
  \caption{We display the behavior of eigenfunctions $\eta(r)$ in the left panels of each rows and $\zeta(r)$ in the right panels of each rows against the radial coordinate $r$ when $\omega=0.4 [10^{-1}km^{-2}]$ corresponding to three variations of $\alpha=1, 0.8, 0.9$.} \label{fig8}
\end{figure}

The set of equations for radial perturbations of our compact stars are modified in this section.
Let us define two parameters by the radial displacement, $\Delta r$, and the pressure perturbation, $\Delta p$ as
\begin{eqnarray}\label{27}
\zeta=\frac{\Delta r}{r}~~~~~\text{and}~~~~~\eta=\frac{\Delta p}{p}.
\end{eqnarray}
which satisfy the following first order differential equations \cite{chanmugam1977radial,vath1992radial}
\begin{eqnarray}\label{28}
\zeta'(r)=-\left(\frac{3}{r} + \frac{p'}{p + \rho}\right)\zeta - \frac{1}{r \Gamma}\eta,
\end{eqnarray}
\begin{eqnarray}\label{29}
\eta'(r)=\tilde{\omega}^{2}\left[r\left(1 + \frac{\rho}{p}\right)\frac{e^{-2\Phi(r)}}{[1-f(r)]}\right]\zeta - \left[\frac{4 p'}{p} + 8\pi(\rho + p)\frac{r}{[1-f(r)]} - \frac{r p'^{2}}{p(\rho + p)}\right]\zeta - \left[\frac{\rho p'}{p(\rho + p)} + 4\pi (\rho + p)\frac{r}{[1 + f(r)]}\right]\eta,
\end{eqnarray}
where $\Gamma$ is the relativistic adiabatic index, defined in the Eq.~(\ref{25}). Here $\tilde{\omega}=s \tilde{\omega}_{0}$ is the frequency oscillation mode with a dimensionless number $s$. Thus, $s=\tilde{\omega}/\tilde{\omega_{0}}$ or, $\nu=s \tilde{\omega_{0}}/(2\pi)$. The constant $\tilde{\omega}_{0}$ may be computed by $\tilde{\omega}_{0}=\sqrt{M/R^{3}}$. \\
The unknown frequencies are computed by solving Strum-Liouville boundary value problem imposing the following boundary conditions, one at the center of the star as $r\rightarrow 0$ and another at the surface $r=R$ \cite{sagun2020asteroseismology}
\begin{eqnarray}
\frac{\eta}{\zeta}\bigg|_{r=0}=-3\Gamma(0)~~~\text{and}~~~\frac{\eta}{\zeta}\bigg|_{r=R}=\left[-4 - \left(1- f(R)\right)^{-1}\left(\frac{f(R)}{2} + \frac{2\tilde{\omega}^{2}R^{2}}{f(R)}\right)\right].
\end{eqnarray}
These conditions are predicted as follows: $\zeta'(r)$ must be finite for $r\rightarrow 0$ and $\eta'(r)$ must be finite at the surface i.e., $\rho,~p \rightarrow 0$. Also, note that $M$ and $R$ are the mass and the total radius of the star respectively. Now, we want to concentrate on some literature in which radial oscillations of compact stellar objects are discussed. In the ref.~\cite{mohanty2024unstable}, authors have explored radial oscillations for unstable anisotropic neutron stars. Radial Oscillations of Quark Stars Admixed with dark matter has been discussed in \cite{jimenez2022radial} and also, radial oscillations of dark matter stars admixed with dark energy has been discussed in \cite{sepulveda2024radial}. Moreover, radial oscillations of neutron stars has been discussed in the refs.~\cite{rather2023radial,sen2023radial,pretel2020equilibrium,sagun2020asteroseismology}. Now, in the Figs.~\ref{fig7} and \ref{fig8}, we have displayed the nature of eigenfunctions $\eta(r)$ and $\zeta(r)$ in the context of radial coordinate $r$ in $[km]$ units. Here, it is notable that we have considered only two values of HL parameters $\omega$ such as $0.9$ and $0.4$ in $[10^{-1}km^{-2}]$ units. In the Table~\ref{table4}, we have computed the values of frequencies in kHz units correspoding to the fixed values of parameters. In each figures, the overtone numbers i.e., the number of zeros of eigenfunctions are represented by $n$. We have considered several excited modes like $n=1, 2, 3, 4, 5, 6$. Here we observe that corresponding to each values of $\omega$ and $\alpha$, the nature $\eta(r)$ and $\zeta(r)$ are very similar for each mode.
%%%%%%%%%%%%%%%%%%%%%%%%%%%%%%%%%%%%%%%%%
%%%%%%%%%%%%%%%%%%%%%%%%%%%%%%%%%%%%%%%%%
%%%%%%%%%%%%%%%%%%%%%%%%%%%%%%%%%%%%%%%%%
%%%%%%%%%%%%%%%%%%%%%%%%%%%%%%%%%%%%%%%%%

%%%%%%%%%%%%%%%%%%%%
\begin{table}
  \centering
  \begin{tabular}{ |p{2cm}|p{2.3cm}|p{2.3cm}|p{2.8cm}|p{2cm}|p{2cm}|p{2cm}| }
 \hline
  \centering $\boldsymbol{\omega [10^{-1}km^{-2}]}$ & \centering $\boldsymbol{n=1}$ & \centering $\boldsymbol{n=2}$ & \centering $\boldsymbol{n=3}$ & \centering $\boldsymbol{n=4}$ & \centering $\boldsymbol{n=5}$  & ~~~$\boldsymbol{n=6}$\\[0.8mm]
 \hline
 \multicolumn{7}{|c|}{\textbf{\textit{\large{$\boldsymbol{\alpha=1}$}} }} \\[1.3mm]
 \hline
  \centering 0.9   &   \centering 3.43819   &  \centering 6.14554  &  \centering 8.61166  &  \centering 11.0054  &\centering 13.3688  & ~~~ 15.7179 \\[1.1mm]
  \hline
  \centering 0.4  & \centering 2.76254   & \centering 4.86663 &  \centering 6.79282 &  \centering 8.6668 & \centering 10.5197 &~~~ 12.3633 \\[1.1mm]
  \hline
  \multicolumn{7}{|c|}{\textbf{\textit{\large{$\boldsymbol{\alpha=0.9}$}} }} \\[1.3mm]
 \hline
  \centering 0.9   &   \centering 3.06223   &  \centering 5.48942  &  \centering 7.6911  &  \centering 9.82443 &\centering 11.9268 & ~~~ 14.0131 \\[1.1mm]
  \hline
  \centering 0.4  & \centering 2.61432   & \centering 4.63362 &  \centering 6.47477 &  \centering 8.26049 & \centering 10.0214 &~~~ 11.7694 \\[1.1mm]
  \hline
  \multicolumn{7}{|c|}{\textbf{\textit{\large{$\boldsymbol{\alpha=0.8}$}} }} \\[1.3mm]
 \hline
  \centering 0.9   &   \centering 2.64522  &  \centering 4.71912  &  \centering 6.61275  &  \centering 8.44762 &\centering 10.2556 & ~~~ 12.0495 \\[1.1mm]
  \hline
  \centering 0.4  & \centering 2.36064   & \centering 4.1959 &  \centering 5.86908 &  \centering 7.49025 & \centering 9.08735 &~~~ 10.6713 \\[1.1mm]
  \hline
\end{tabular}
  \caption{We reported frequencies of radial oscillation modes corresponding to values of $\omega$ like $0.9,~0.4 ~[10^{-1}km^{-2}]$ in each case of $\alpha=1,~0.9,~0.8$.}\label{table4}
\end{table}
%%%%%%%%%%%%%%%%%%%%%%%%%%%%%%%%%%%

\section{Conclusions}\label{sec5}

In this paper, we have investigated the basic physical properties of spherical, non-rotating configurations made of isotropic matter in the context of HL gravity. In particular, we have considered the interior matter of stellar system obeys more extended Chaplygin gas EoS (\ref{13}), a specific relation between density and pressure, and hence named by dark energy star. Utilizing this EoS, we numerically solved the  modified TOV equation with proper boundary conditions in HL gravity and obtained final output for mass $m$, energy density $\rho$ and pressure $p$ as a function of radial coordinate $r$. In this study, our main goal is to investigate the deviation of HL gravity from GR in the effects of a single parameter $\omega$. Particularly, we have emphasized on mass-radius relation with estimation of the maximum mass $M_{mas}$ and corresponding radius $R_{predicted}$ and hence compactness. Further, we have discussed the stability of the obtained stellar structure in HL gravity. \\
This study started with the notion of relevant testing of HL gravity in an extreme environments such as dark energy stars. All results regarding the changes of physical properties for dark energy stars have showed the sensitivity of the model parameter $\omega$ through graphical analysis. The numerical estimation of maximum mass and corresponding predicted radius, reported in Tables~\ref{table1},~\ref{table2}, and~\ref{table3}, for three consequent values of $\alpha$ like $1,~0.9,~0.8$ gradually decreases while $\omega$ increases. By observing the Fig.~\ref{fig4}, one can also understand the same results that the curve lines are gradually decreases corresponding to each increase value of $\omega$. Further, we have shown when $\omega\rightarrow \infty$ our obtained results are coincide with that of GR. Subsequently, we have compared our obtained results with some observed pulsar like compact stars data. Such as, it is predicted that this model may suitable to describe the massive pulsars like $PSR J0952-0607$, $PSR J0740+6620$, and $PSR J0751+1807$ in HL gravity. Furthermore, we have calculated maximum compactness $(M/R)_{max}$ and surface redshift $(Z_{sur})_{max}$, given in the Tables, corresponding to the maximum mass $M_{max}$. The nature of them are graphically represented through the respective Figs.~\ref{fig4a} and \ref{fig4b}. As a conclusion, we have the present dark energy stars model satisfy phsically viability conditions. Most importantly, we have calculated the predicted Schwarzschild radius $R_{Sch}$ for each variation of $\omega$ and observed that $R_{Sch}<R_{predicted}$ in each case of $\alpha=1, ~0.9, ~0.8$. Thus, our proposed dark energy stellar system never collapse into a black hole. \\
Finally, we analyzed stability of obtained stellar system via sound speed $c_{s}$, adiabatic index $\Gamma$ and static stability criteria. Our calculations showed that considered dark energy stars model satisfy stability conditions like $0<c_{s}^{2}<1$ and $\Gamma>4/3$ in each variations of model parameters $\omega$ and $\alpha$ in HL gravity. Also, we have shown that $M_{max}$ is a turning point from stability to instability. As a results, we have our considered dark energy stars structure is stable against infinitesimal radial oscillations and sound propagation. \\
Next, we have computed the frequencies and nature of eigenfunctions $\eta(r)$ and $\zeta(r)$ against $r$ by integrating the perturbations equations in support of appropriate boundary conditions at the center and surface of the stellar object. For more details see the Table~\ref{table4} and Figs.~\ref{fig7}, \ref{fig8}. Here, we have considered only six lowest excited modes corresponding to two HL parameters like $\omega=0.9, 0.4$. It is observed that frequencies are increasing when $\omega$ increases. \\
 
Overall, the HL gravity provides a good result in analyzing of mass-radius relation, stability and some other features for the exploration of dark energy stars with a specific extended Chaplygin gas EoS. It is well known that compact stars including dark energy stars with mass grater than $2M_{\odot}$ is very difficult to evaluate in the framework of standard GR. However, in HL gravity, we can rescue this problem for dark energy stars. Additionally, our present stellar system stable through several perspective. So, this model represents a physically acceptable and viable stable dark energy stellar structure with heavier mass in HL gravity than in GR.\\

\textbf{\Large{Acknowledgements:}} KPD is thankful to UGC, Govt. of India,for providing Senior Research Fellowship (F.NO.16-6(DEC.2017)/2018(NET/CSIR)).\\

\section*{Data Availability}
Our present article is developed theoretically and all fruitful results are constructed from stated equations. We did not generate no new data during this study. Thus, present article has no associate data. \\

\bibliographystyle{naturemag}

\bibliography{bibliography}

\end{document}